\documentclass[
reprint,
amsmath,
amssymb,
floatfix,
superscriptaddress,
aps,
prr
]{revtex4-2}
\usepackage{graphicx}
\usepackage{dcolumn}
\usepackage{bm}
\usepackage{color}
\usepackage{amssymb}
\usepackage{amsmath}
\usepackage{xfrac}
\usepackage{mathrsfs}
\usepackage{comment}

\usepackage[normalem]{ulem}

\usepackage{hyperref}
\usepackage[dvipsnames]{xcolor}
\hypersetup{
    colorlinks=true,
    linkcolor=Blue,
    citecolor=Blue,   
   urlcolor=Blue
   }

\def\bea{\begin{eqnarray}}
\def\eea{\end{eqnarray}}
\def\beq{\begin{equation}}
\def\eeq{\end{equation}}
\def\f{\frac}

\def\la{\langle}
\def\ra{\rangle}
\def\nn{\nonumber}

\def\d{\delta}
\def\p{\partial}

\def\W{\Omega}

\def\uv{ {\hat{\mathbf{u}}}}
\def\rv{ {\bf r}}

\def\d{\delta}
\def\p{\partial} 

\def\la{\langle}
\def\ra{\rangle}

\begin{document}

\title{Steering chiral active Brownian motion via stochastic position–orientation resetting}

\author{Amir Shee}%
\email[]{amir.shee@uvm.edu}
\affiliation{Department of Physics, University of Vermont, Burlington, VT 05405, United States of America}

\vspace{10pt}

\begin{abstract}
Guiding active motion is important for targeted delivery, sensing, and search tasks.
Many active systems exhibit circular swimming, ubiquitous in chemical, physical, and biological systems, that biases motion and reduces transport efficiency.
We show that stochastic position–orientation resetting can overcome these limitations in two-dimensional chiral active Brownian particles by interrupting circular motion, resulting in tunable dynamics.
When resets are infrequent compared to chiral rotation, the steady-state mean-squared displacement varies non-monotonically with rotational diffusion.
Steady state excess kurtosis and orientation autocorrelation yields spatiotemporal state diagram comprising three states: an activity-dominated chiral state, and two resetting-dominated states with and without chiral rotation; in contrast, the achiral(or non-chiral) counterpart supports only the resetting-dominated state without chiral rotation.
Chirality thus enriches the dynamical landscape, enabling tunable transitions between transport modes absent in achiral systems.
A simple reset protocol can therefore transform chiral active dynamics and offer a practical strategy for optimizing search and transport in circle swimmers.
\end{abstract}

\maketitle

\section{Introduction}

Active matter comprises self-propelled particles that continuously convert energy into directed motion, making it a paradigmatic example of a non-equilibrium system~\cite{Marchetti2013, Bechinger2016}. 
In particular, active Brownian particle (ABP) represent minimal model that capture the competition between persistent self-propulsion and stochastic reorientation driven by noise~\cite{Romanczuk2012, Bechinger2016, Basu2018, Shee2020, Baconnier2025}. 
Active Brownian particle with an intrinsic angular velocity (i.e., chirality), termed chiral active Brownian particle (in short referred to as chiral ABP, cABP, or CABP in the literature), constitute a distinct subclass that exhibits well defined circular trajectories~\cite{Sevilla2016, Caprini2023, Liebchen2017, Liebchen2022, Chan2024, Yang2024, Semwal2024, Cruz2024, Shee2024, Pattanayak2024}.
The interplay among self-propulsion, chirality, and rotational diffusion gives rise to rich dynamical phenomena such as circular, spiral trajectories, enhanced diffusion under optimal conditions~\cite{Pattanayak2024, Sevilla2016, Liebchen2017, Caprini2024}.
Chiral active motion has been observed experimentally in diverse systems, including active biomolecules such as proteins, cytoskeletal microtubules, unicellular bacteria, and motile sperm cells~\cite{Loose2014, Sumino2012, Brokaw1982, DiLuzio2005, Lauga2006, Riedel2005, Nosrati2015}.
These unique dynamics motivate the exploration of control mechanisms, such as resetting, to manipulate their behavior.


Stochastic resetting is a protocol in which a system’s dynamics are intermittently interrupted by returning it to a predetermined state, thereby introducing an external time scale that competes with the intrinsic dynamics~\cite{Evans2011, Majumdar2015, Evans2020, Tal-Friedman2020}.
This procedure not only alters the transient behavior but also leads to non-equilibrium steady states (NESS) with statistical properties that differ markedly from those of equilibrium systems governed by detailed balance.
For instance, in diffusion-limited search processes, resetting prevents the searcher from wandering indefinitely, optimizing the mean first-passage time and offering practical advantages in algorithm design, ergodicity, ecological foraging, and financial modeling~\cite{Luby1993, Pal2014, Reuveni2014, Pal2019, Wang2021, Vinod2022, GuptaJayannavar2022, Wang2022, Santra2022, NagarGupta2023, Biswas2024, Liang2025}.
These effects highlight resetting as a powerful tool for controlling nonequilibrium systems and demonstrate its applicability to more complex dynamics such as in active matter~\cite{Kumar2020}.

Recent advances have demonstrated that stochastic resetting can be harnessed as a control mechanism to improved search efficiency and modified transport properties~\cite{Evans2011, Reuveni2014, Majumdar2015}.
In active matter, resetting has been employed to model phenomena such as bacterial tumbling~\cite{Berg2004} and to regulate the behavior of self-propelled particles~\cite{Kumar2020, Sar2023, Baouche2024, Parmanick2024, Olsen2024, Shee2025, Baouche2025, Olsen2025}.
The chiral active particle with strategic reorientation has recently been investigated in~\cite{Olsen2024}.
However, the combined effects of chirality and stochastic resetting on active particle dynamics remain unknown.
Understanding the influence of resetting on circular motion will drive future strategies for controlling transport in chiral active systems.
Here, we develop a theoretical framework that systematically reveals how externally imposed resets compete with intrinsic chiral rotation to regulate transport statistics, providing a physically grounded route to steering, optimizing exploration, and tuning dynamical regimes in chiral active matter.

Chiral active particles often move along circular trajectories that constrain their spatial exploration and diminish transport efficiency, as observed in bacteria and sperm cells~\cite{DiLuzio2005, Riedel2005, Lauga2006, Nosrati2015}.
Similar chiral rotation is also found in engineered systems, including asymmetric active colloids~\cite{Kummel2013}, micro-rotors and confined spinners~\cite{Lowen2016, LopezCastano2022}, and actively spinning micro-vesicles~\cite{Kaur2025}.
These systems provide natural experimental platforms for testing our predictions under an experimentally accessible external strategy to interrupt active chiral rotation such as stochastic resetting~\cite{Tal-Friedman2020}.
Other controlled interruptions can be realized experimentally using optical or magnetic tweezers~\cite{Grzybowski2000, Yan2015, Tal-Friedman2020}, intermittent illumination of light-activated Janus particles~\cite{Palacci2013, Golestanian2007, Mano2017}, or activation–deactivation cycles in chemically powered nanomotors~\cite{Mirkovic2010}.
Moreover, our predictions can be tested in macroscopic robotic active-matter experiments using Hexbug robots~\cite{Dauchot2019} operated under stochastic resetting protocols~\cite{Pramanick2024, Olsen2025}.

\begin{figure}[!t]
\begin{center}
\includegraphics[width=0.5\linewidth]{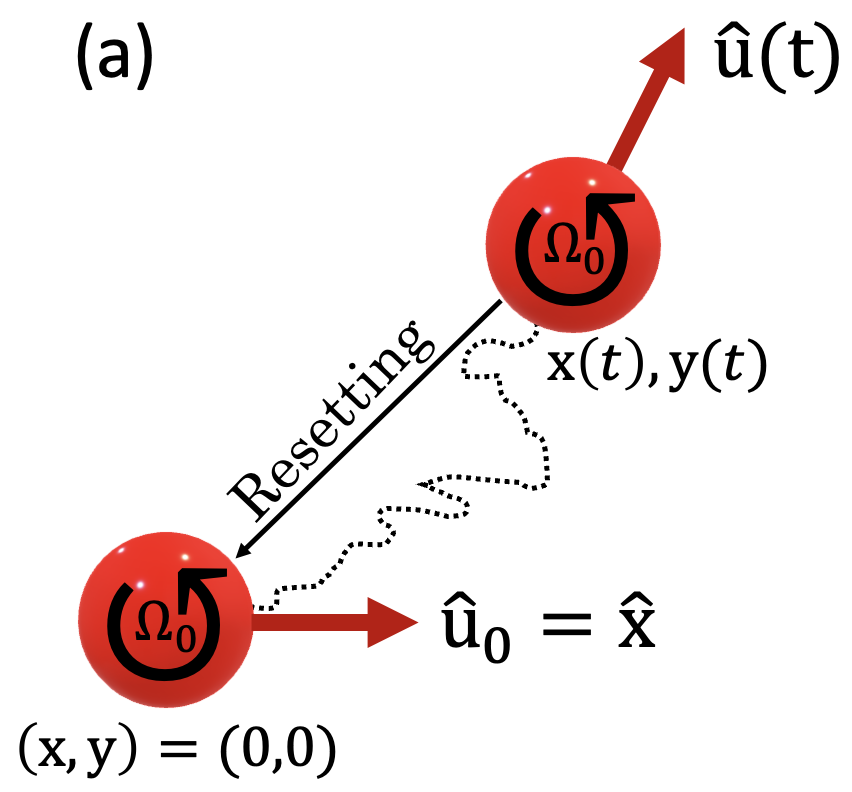} 
\includegraphics[width=\linewidth]{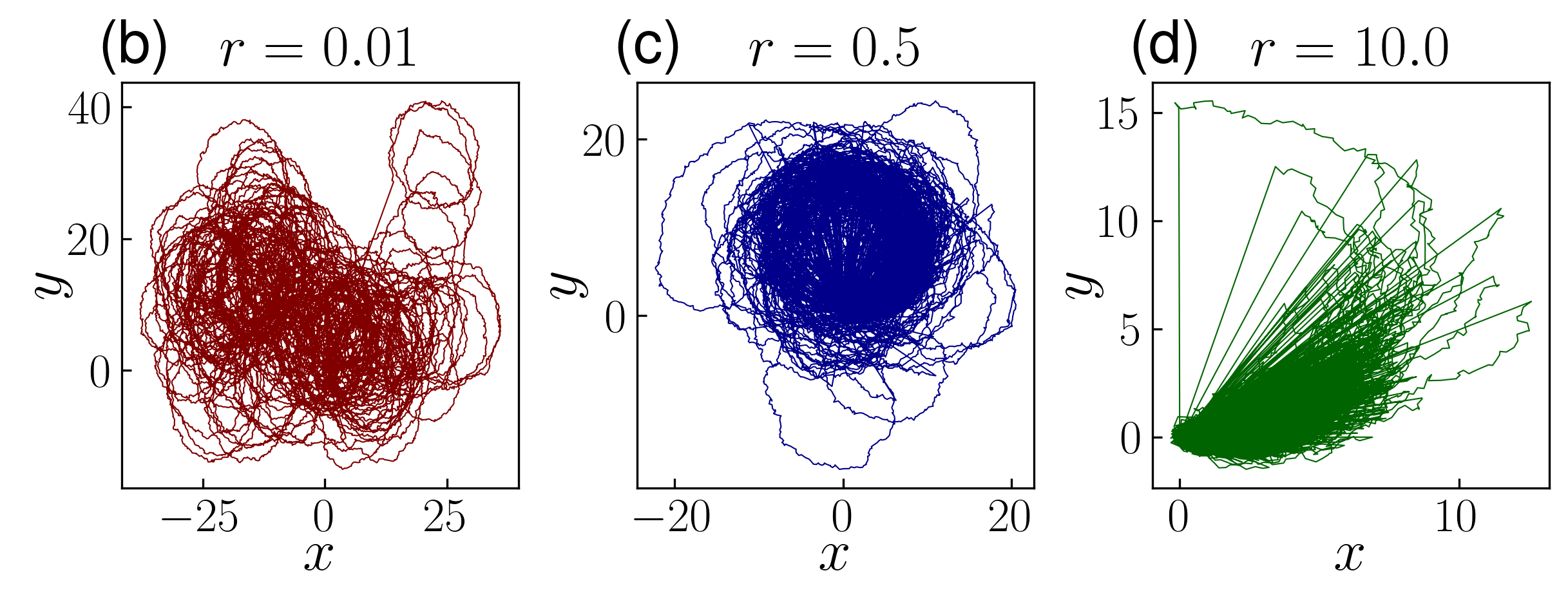} 
\caption{
Schematic representation of the dynamics of a chiral active Brownian particle subject to stochastic resetting of both its position and orientation (a).
Steady-state trajectories for reset rates $r=0.01$ (b), $0.5$ (c), and $10$ (d), obtained using parameters $v_{0}=20$, $\Omega_{0}=2.5$, $D=1$, and $D_{r}=0.05$. Increasing $r$ progressively suppresses spatial exploration from (b,c) chiral rotation($r+D_r<\W_0$) with (b) scattered loop and (c) concentrated loop to (d) non-rotating($r+D_r>\W_0$) excursions.
} 
\label{fig1}
\end{center}
\end{figure}

In this paper, we study a chiral active Brownian particle in two dimensions~\cite{Sevilla2016, Liebchen2022, Pattanayak2024} under stochastic resetting by implementing a protocol that resets both its position and orientation to fixed initial values at a constant rate~\cite{Kumar2020, Shee2025}.
This resetting competes with both the circular motion induced by chirality and the randomizing effects of rotational diffusion, giving rise to spatiotemporal dynamics.
We employ a renewal‐equation framework together with a Fokker–Planck formalism~\cite{Shee2020, Shee2025} to derive exact analytical expressions for the moments to characterize its steady‐state distributions.
Our analytical predictions are corroborated by numerical simulations, which reveal interplay among resetting, self-propulsion, and chirality.

The remainder of this paper is organized as follows. In Sec.~\ref{sec:model}, We introduce the model for a chiral active Brownian particle under stochastic resetting and outline our theoretical framework. In Sec.~\ref{sec:lower_order_moments}, we present results for lower-order moments, including orientation dynamics and the mean-squared displacement (MSD). Sec.~\ref{sec:state_diagrams} examines fourth-order moments and excess kurtosis, and introduces a state diagram identifying activity-dominated and resetting-dominated regimes. Finally, Sec.\ref{sec:conclusions} discusses the implications of our findings for controlling chiral active systems and potential applications.

\section{Model}
\label{sec:model}

A chiral active Brownian particle(CABP) in two dimensions is described by its position $\mathbf{r} = (x, y)$, and its orientation (or heading direction) $\uv = (u_x, u_y)$, which is a unit vector in two dimensional plane where $u_x=\cos(\theta)$ and $u_y=\sin(\theta)$, evolving over time $t$ in the presence of translational diffusion coefficient $D$. The position $\mathbf{r}$ evolves with constant active speed $v_0$ along active orientation direction $\uv$ evolving with constant angular velocity i.e., chirality $\W_0$ with orientation fluctuation sets by orientation diffusion coefficient $D_r$. Stochastic resetting imposed on both $\rv$ and $\uv$ intermittently resets to initial values $(\rv_0,\uv_0)$ with rate $r$. The position $\rv$ and orientation $\uv$ evolves 
\bea
&&\dot \rv = v_0 \uv + \sqrt{2D} \boldsymbol{\chi}(t)\,,
\label{eom:disp}\\
&&\dot \uv =  [\W_0 + \sqrt{2D_r}\eta (t)]\uv^{\perp}\,,
\label{eom:rot_active}\\
&&(\rv, \uv) \rightarrow (\rv_0, \uv_0)\,.
\label{eom:resetting}
\eea
$\uv^{\perp} = (-u_y, u_x)$ is the unit vector perpendicular to $\uv$.
The dynamics is illustrated schematically in Fig.~\ref{fig1}(a).
The noise terms $\boldsymbol{\chi}(t)$ and $\eta(t)$ are modeled as Gaussian white noise with zero mean and variances given by $\la \chi_i(t) \chi_j(t^{\prime})\ra = \d_{ij} \d(t-t^{\prime})$ and $\la\eta(t)\eta(t^{\prime})\ra=\d(t-t^{\prime})$, respectively.
The interplay of the three timescales $\W_0^{-1}$, $D_{r}^{-1}$, and $r^{-1}$ determines the steady‐state behavior of the chiral ABP under resetting.

We initialize the particle at the origin with its orientation along the $x$ axis, and apply reset to the initial state $(x,y,\theta)=(0,0,0)$ at rate $r$.
We numerically integrate Eqs.~\eqref{eom:disp} and \eqref{eom:rot_active} using the Euler–Maruyama scheme, incorporating stochastic resets via Eq.~\eqref{eom:resetting}. Steady-state trajectories are then used to compute numerical moments, which we compare directly with the corresponding analytical steady-state predictions presented below.
Figure~\ref{fig1}(b-d) visualizes the steady-state trajectories corresponding to weak (b), intermediate (c), and strong (d) resetting.

\medskip
\noindent
{\bf Moments generator framework:}
The probability distribution $P(\rv, \uv, t)$ of the position $\rv$ and the active orientation $\uv$ of the particle without stochastic resetting follows the Fokker-Planck equation~\cite{Hermans1952, Shee2020}
\bea
\p_t P(\rv, \uv, t) &=& D\nabla^2 P +D_r \nabla_\uv^2 P - v_0\, \uv\cdot \nabla P \nonumber\\
&&- \W_0 \uv^{\perp} \cdot \nabla_{\uv} P~,
\label{eq:F-P}
\eea
where $\nabla$ and $\nabla_\uv$ are the gradient in spatial and orientation space, respectively.
The right-hand side contains (i) translational diffusion $D\nabla^2 P$, (ii) rotational diffusion on the unit circle $D_r\nabla_{\uv}^2 P$ (here in 2d, $\nabla_{\uv}^2=\partial_\theta^2$), (iii) self-propulsive advection in real space $-v_0\,\uv \cdot \nabla P$, and (iv) deterministic chiral rotation in 2d (advection in orientation space) $-\Omega_0\,\uv^\perp \cdot \nabla_{\uv} P$ (i.e., $-\Omega_0 \partial_\theta P$).

Utilizing the Laplace transform $\tilde P(\rv, \uv, s) = \int_0^\infty dt\, e^{-s t}\, P(\rv, \uv, t) $ and defining the mean of an observable $\la \psi \ra_s = \int d\rv \, d\uv\, \psi(\rv, \uv ) \tilde P(\rv, \uv, s)$, multiplying by $\psi(\rv, \uv)$ and integrating over all possible $(\rv, \uv)$, we find,
\bea
-\la \psi \ra_0 + s \la \psi \ra_s &=& D\la \nabla^2 \psi \ra_s + D_r\la \nabla_\uv^2 \psi \ra_s \nonumber\\
&&+ v_0\la\, \uv\cdot \nabla \psi \ra_s+\W_0 \la \uv^{\perp}\cdot \nabla_{\uv} \psi\ra_s~,
\label{eq:moment}
\eea
where the initial condition sets $\la \psi \ra_0 = \int d\rv \, d\uv\,  \psi(\rv, \uv) P(\rv, \uv, 0)$. Without any loss of generality, we consider the initial condition to follow $P(\rv, \uv, 0) = \d(\rv-\rv_0) \d(\uv - \uv_0)$, where $\rv_0$ and $\uv_0$ are the initial position and orientation respectively. Equation~(\ref{eq:moment}) can utilize to compute exact moments as a function of time without stochastic resetting.

Under stochastic resetting of both position and orientation, the moments satisfy the renewal equation~\cite{Kumar2020, Shee2025}
\bea
\la \psi(t)\ra_r &=& e^{-r t} \la\psi(t)\ra + r \int_{0}^{t} dt^{\prime} e^{-rt^{\prime}} \la\psi (t^{\prime})\ra\,.
\label{eq:moment_resetting}
\eea
This framework yields exact dynamical moments, and the non-equilibrium steady state is obtained from their long-time limit, $\la \psi\ra_r^{\rm st}=\lim_{t\to\infty}\la \psi(t)\ra_r$.
Alternatively, the steady state under resetting can be obtained directly via the Final Value Theorem (FVT) applied to Eq.~\eqref{eq:moment}
\bea
\la\psi\ra^{\rm st}_{r} &=& r \la\psi\ra_s(s=r)\,.
\label{eq:moment_resetting_FVT}
\eea
In the following, we present our main results, while detailed derivations are provided in the Appendix.

\section{Orientation autocorrelation and mean-squared displacement(MSD)}
\label{sec:lower_order_moments}
\noindent
{\bf Orientation autocorrelation:~}
Since the orientation evolves independently of position, its mean value equals its autocorrelation, and with damping arising from resetting, this quantity defines the steady-state autocorrelation.
The orientation autocorrelation calculated using moments generator framework described above (see detailed derivation in Appendix~\ref{app:lower_order_moments}) results
\bea
\la\uv(\tau)\cdot\uv(0)\ra_r &=& e^{-(r+D_r) \tau} \cos(\W_0 \tau)\nonumber\\
&&+ \f{r(r+D_r)[1-e^{-(r+D_r)\tau}\cos(\W_0 \tau)]}{(r+D_r)^2+\W_0^2}\nonumber\\
&&+ \f{r \W_0 e^{-(r+D_r)\tau} \sin(\W_0 \tau)}{(r+D_r)^2+\W_0^2}\,.
\label{eq:ncorr_resetting}
\eea
If $r+D_r< \W_0$ the autocorrelation exhibits damped oscillations; if $r+D_r>\W_0$, it decays monotonically. 
If $r$ is set to zero then Eq~\eqref{eq:ncorr_resetting} reduces to $\la\uv(\tau)\cdot\uv(0)\ra=e^{-D_r \tau} \cos(\W_0 \tau)$ recovering the standard chiral ABP result~\cite{Pattanayak2024}. In the large timescale $\tau\to\infty$, the orientation autocorrelation saturates to finite value
\bea
C &=& \f{r(r+D_r)}{(r+D_r)^2+\W_0^2}\,.
\label{eq:ucorr_st}
\eea
As chirality $\W_0$ increases, the particle rotates more rapidly between resets and explores a broader range of orientations.
Consequently, it loses memory of its initial heading, and decreases with $\W_0$.
In the achiral(or non-chiral) limit $\W_0=0$, Eq.~\eqref{eq:ucorr_st} simplifies to $C= r/(r+D_r)$ for ABP under stochastic resetting~\cite{Shee2025}. Conversely, in the absence of rotational diffusion $D_r=0$, Eq.~\eqref{eq:ucorr_st} becomes  $C= r^2/(r^2+\W_0^2)$, so increasing $r$ yields a steeper recovery of orientation memory compared to the achiral case.

\begin{figure}[!t]
\begin{center}
\includegraphics[width=\linewidth]{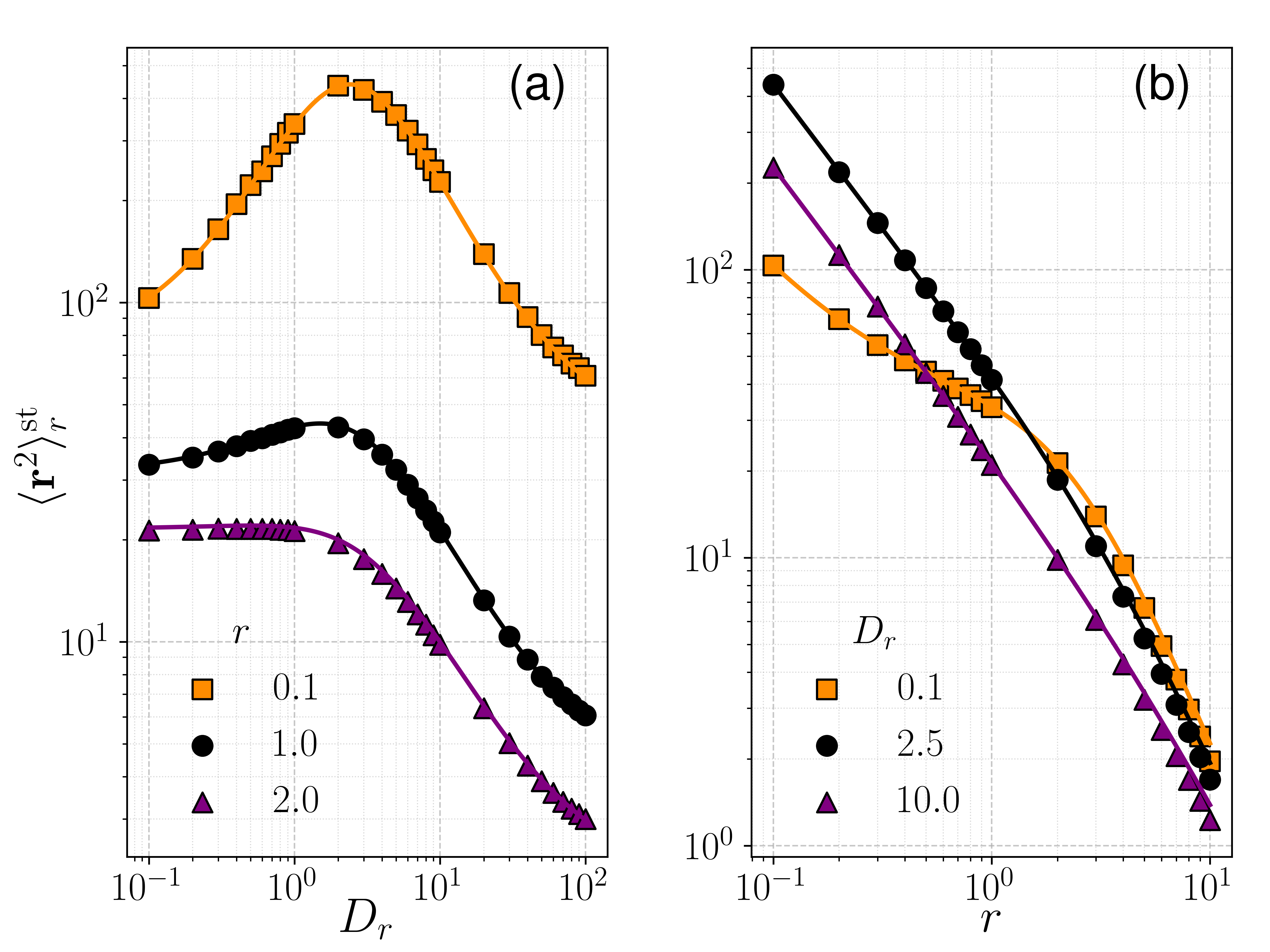} 
\caption{
Steady-state mean squared displacement $\la\rv^2\ra^{\rm st}_{r}$ as a function of $D_r$ in (a) for $r=0.1,~1,~2$ and of $r$ in (b) for $D_r=0.1,~2.5,~10$.
Resetting suppressed non-monotonic behavior of $\la\rv^2\ra^{\rm st}_{r}$ with $D_r$.
Solid lines are analytic prediction of Eq.~\eqref{eq:r2avg_reset_st} and symbols are from simulation.
Initial position is at the origin with the initial orientation along the $x$-axis.
Fixed parameters are $v_0=10$, $\W_0=2.5$, and $D=1$.
} 
\label{fig2}
\end{center}
\end{figure}

\medskip
\noindent
{\bf Mean-squared displacement(MSD):~}
In the long time limit $(t\to \infty)$, mean-squared displacement (MSD) reaches to steady state as $ \la\rv^2\ra_{r}^{\rm st}=\la\rv^2\ra_r(t\to\infty)$,
\bea
\la\rv^2\ra_{r}^{\rm st} &=& \f{4D}{r} + \f{2 (r+D_r)  v_0^2 }{r((r+D_r)^2+\W_0^2)}\,.
\label{eq:r2avg_reset_st}
\eea
Appendix~\ref{app:MSD} presents the derivation of Eq.~\eqref{eq:r2avg_reset_st}.
The first term, $4D/r$, arises purely from translational diffusion punctuated by resets: more frequent resets(large $r$) confine the diffusive spread.
The second term captures the active contribution: the self-propulsion speed $v_0$ drives exploration, but its effectiveness is reduced both by rotational noise($D_r$) and by chiral rotation($\W_0$), with resets competing against these decorrelating processes.
Figure~\ref{fig2} compares the analytic prediction of the steady‐state MSD (solid lines, Eq.~\eqref{eq:r2avg_reset_st}) with numerical results (symbols), demonstrating excellent agreement.
Maximizing $\la\rv^2\ra_{r}^{\rm st}$ with respect to $D_r$ holding $r$, $v_0$, and $\Omega_0$ fixed; gives optimal rotation diffusion $D_r^{*}=\mathrm{max}\{\Omega_0-r, 0\}$; so that $\Omega_0> r$ optimum occurs at $r+D_r = \Omega_0$ and $(\la\rv^2\ra_{r}^{\rm st})_{\rm max}=[4D + v_0^2/\Omega_0]/r$.
Physically, this corresponds to matching the total reorientation rate (rotational diffusion plus resets) to the intrinsic spin rate, thereby maximizing net displacement before being reset.
The non-monotonic dependence in Fig.~\ref{fig2}(a) has no analogue in the achiral resetting case~\cite{Shee2025} and reflects a new result unique to chiral active motion.
The maximization of MSD with rotational noise qualitatively corresponds to the strategic reorientation~\cite{Olsen2024} and \emph{hammering state} observed in densely packed interacting chiral ABPs, or equivalently in a single chiral ABP confined in a harmonic trap~\cite{Debets2023, Shee2024}.
If $\Omega_0\leq r$, no finite $D_r$ can beat the resets, and the optimum lies at $D_r=0$. 
In the noiseless limit by setting both translational and rotational noise zero $D=D_r=0$, steady-state MSD in Eq.~\eqref{eq:r2avg_reset_st} simplifies to $\la\rv^2\ra_{r}^{\rm st} = 2 v_0^2 / (r^2+\W_0^2)$, decays once the reset rate or chirality dominates.

\begin{figure}[!t]
\begin{center}
\includegraphics[width=\linewidth]{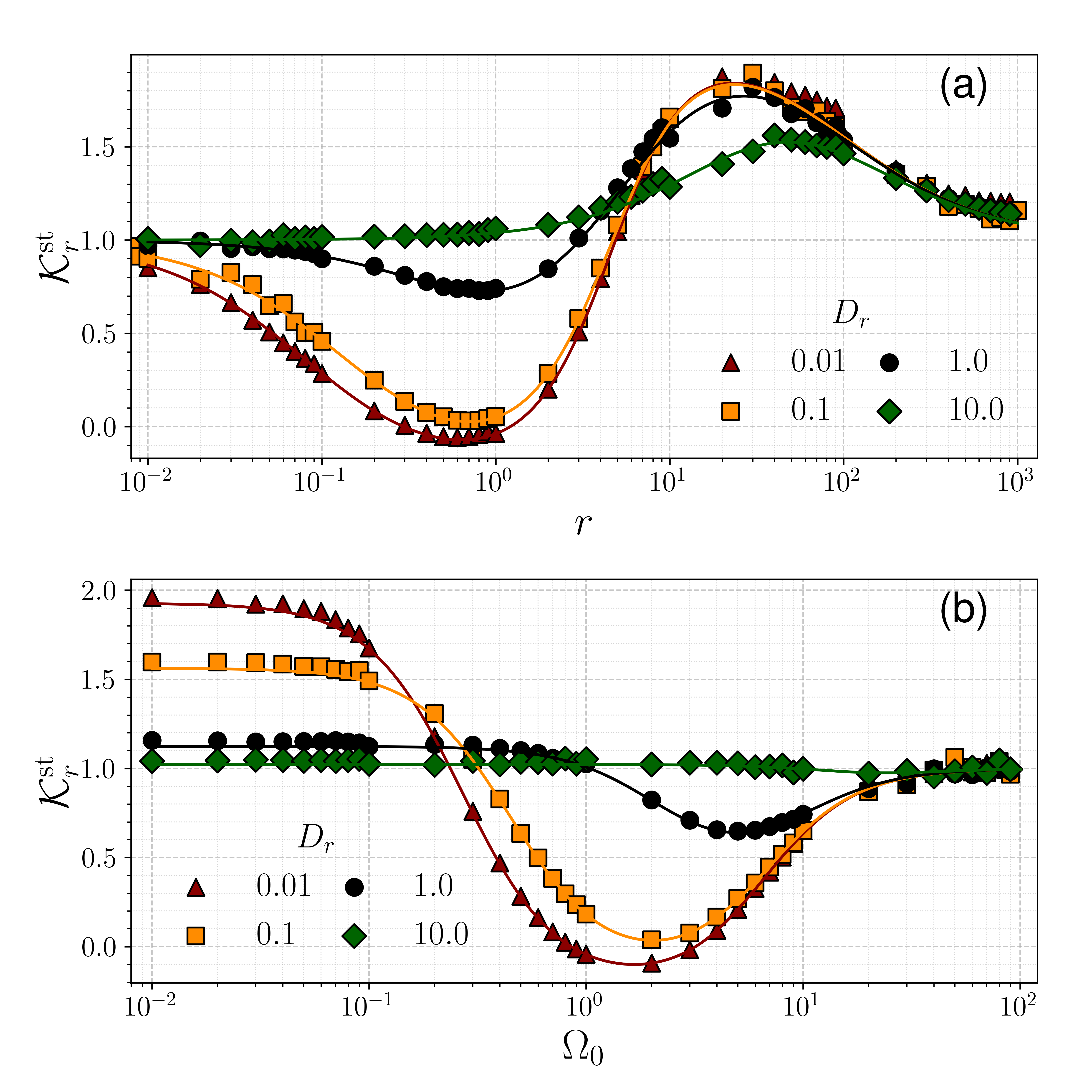} 
\caption{
Steady-state excess kurtosis $\mathcal{K}_{r}^{\rm st}$ as function of $r$ in (a) and of $\W_0$ in (b) for rotational diffusion coefficient $D_r=0.01,~0.1,~1,~10$.
Negative values of $\mathcal{K}_{r}^{\rm st}$ indicate the weakly active state in both plots.
Symbols correspond to simulation, and lines show the prediction from Eq.~\eqref{eq:excess_kurtosis_st}.
Fixed parameters are $v_0=10$, $\W_0=2.5$, and $D=1$.
} 
\label{fig3}
\end{center}
\end{figure}

\medskip
In Fig.~\ref{fig2}(a), the maximum of $\la\rv^2\ra_{r}^{\rm st}$ occurs at $D_r^{*}=\Omega_0-r=2.4, 1.5, 0.5$ for $r=0.1, 1.0, 2.0$ respectively, with $\Omega_0=2.5$.
The maximum of the steady‐state MSD for $D_r>0$ vanishes at the reset rate $r = \Omega_0$.
As the reset rate approaches $\Omega_0$, resetting alone suffices to counteract chirality, so no additional rotational noise can further enhance the MSD and the peak vanishes.
In the small-$r$ limit, a moderate amount of rotational diffusion enables the particle to escape its looping trajectories, giving rise to the reentrant increase in $\langle \mathbf{r}^{2} \rangle_{r}^{\mathrm{st}}$ with $D_{r}$; by contrast, at large $r$, the motion is dominated by frequent resets ($\langle \mathbf{r}^{2} \rangle_{r}^{\mathrm{st}} \sim 4D/r$), and additional rotational noise only suppresses the overall spread, leading to a monotonic decrease of $\langle \mathbf{r}^{2} \rangle_{r}^{\mathrm{st}}$ with $D_{r}$ [Fig.~\ref{fig2}(b)].

\medskip
The effective diffusion coefficient in the steady state is calculated as $D_{\rm eff}= r \la\rv^2\ra_{r}^{\rm st}/4$, which characterizes the effective spatial spread of the particle in the nonequilibrium steady state induced by stochastic resetting. It follows that
\bea
D_{\rm eff} &=& D + \f{ (r+D_r)  v_0^2 }{2((r+D_r)^2+\W_0^2)}\,.
\label{eq:Dex}
\eea
We define excess diffusion coefficient $D_{\rm ex} = D_{\rm eff} - D$, which quantifies the deviation from the translational diffusion coefficient.
Physically, $D_{\mathrm{ex}}$ measures the additional enhancement of spatial exploration arising from the particle’s self-propulsion and its interplay with stochastic resetting and chirality, beyond the passive Brownian contribution $D$.
In the absence of both stochastic resetting ($r = 0$) and chirality ($\W_{0} = 0$), the excess diffusion coefficient becomes $D_{\mathrm{ex}} = v_{0}^{2}/2D_{r}$, representing the excess diffusion coefficient for ABP.
In the absence of chirality ($\W_{0}=0$), the excess diffusion coefficient becomes $D_{\mathrm{ex}} = v_{0}^{2}/2(r + D_{r})$, corresponding to an ABP under stochastic position-orientation resetting~\cite{Shee2025}. In the absence of stochastic resetting ($r = 0$), the excess diffusion coefficient reduces to $D_{\mathrm{ex}} = D_{r} v_{0}^{2} / 2(D_{r}^{2} + \W_{0}^{2})$ for a CABP~\cite{Pattanayak2024}.

\begin{figure*}[!t]
\begin{center}
\includegraphics[width=0.48\linewidth]{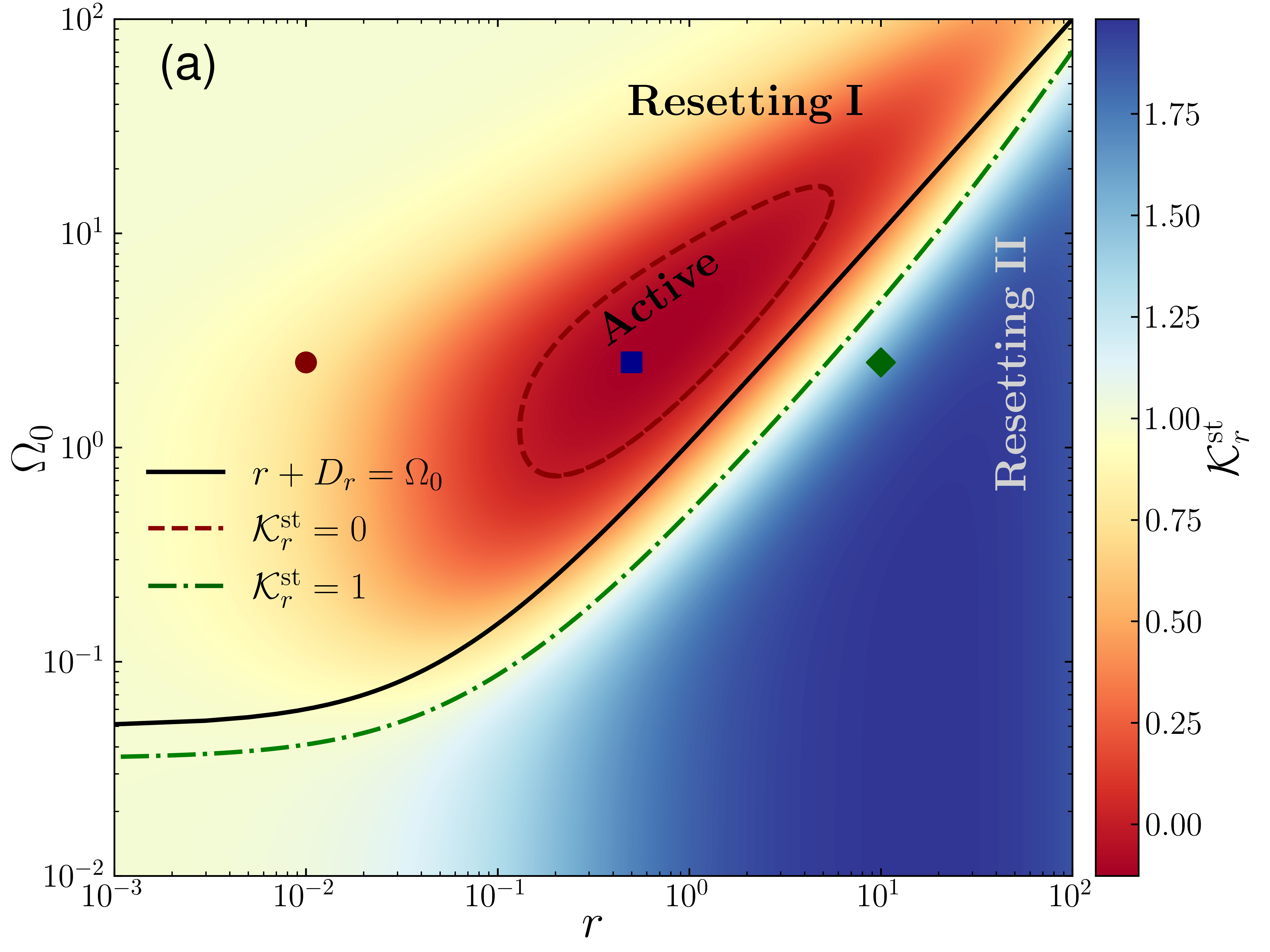} 
\includegraphics[width=0.48\linewidth]{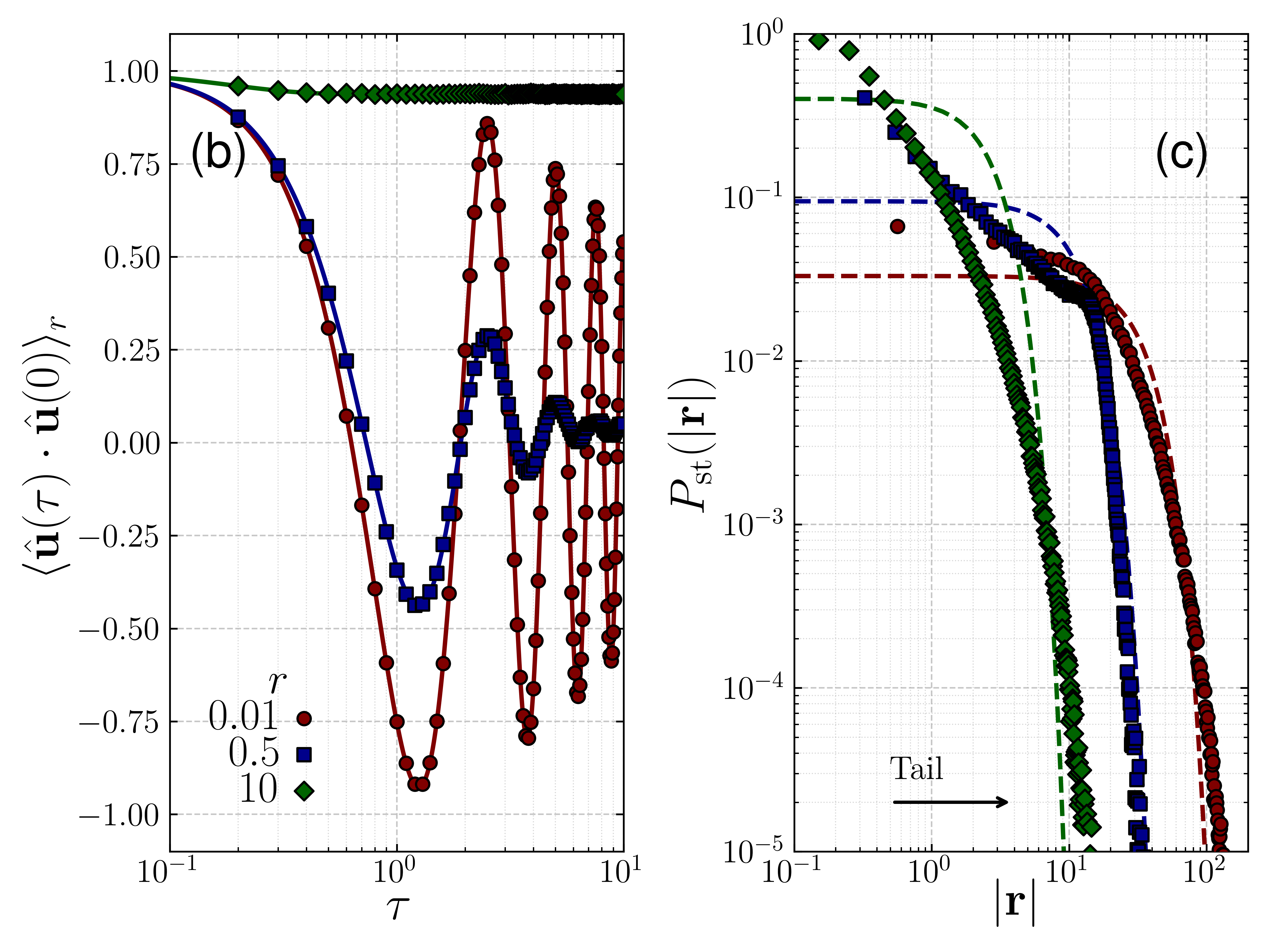} 
\includegraphics[width=\linewidth]{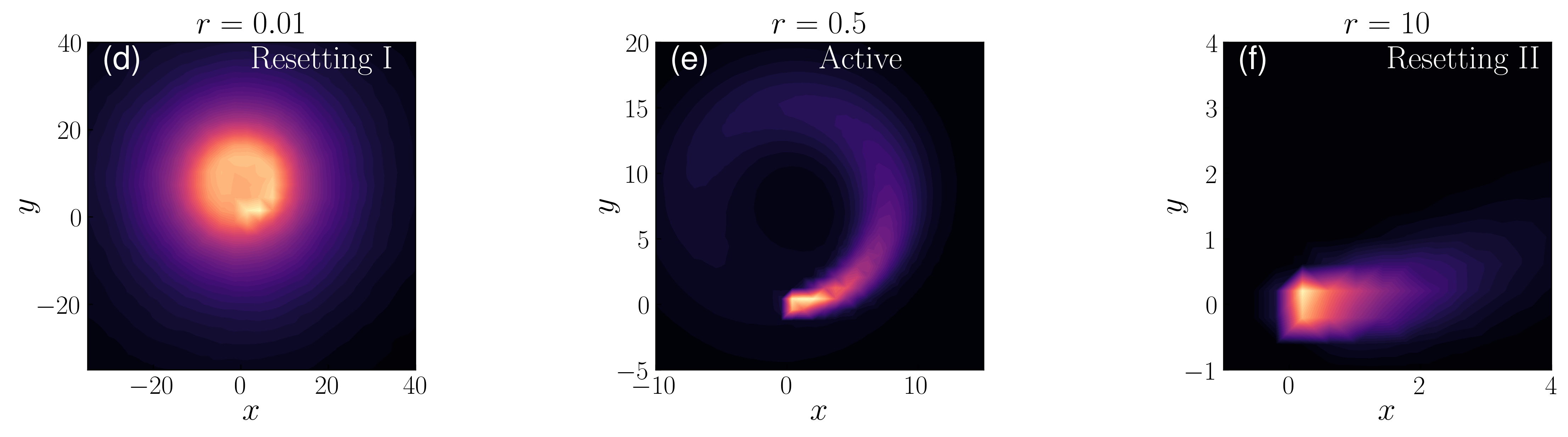}
\caption{
(a) State diagram on $r-\W_0$ plane. The colormap depicts the steady-state excess kurtosis $\mathcal{K}_{r}^{\mathrm{st}}$ (Eq.~\eqref{eq:excess_kurtosis_st}). The black solid line correspond to $r+D_r=\W_0$ sets boundary between oscillatory (chiral rotation) and non-oscillatory (Non-rotating) regime. The red dashed line correspond to passive state by $\mathcal{K}^{\rm st}_{r}=0$ sets boundary between resetting ($\mathcal{K}^{\rm st}_{r}>0$) or activity ($\mathcal{K}^{\rm st}_{r}<0$) dominated regime. Three regions identified: Rotating Active, rotating Resetting, and Non-Rotating Resetting while Non-Rotating Active ($\mathcal{K}^{\rm st}_{r}<0$ with $\W_0 < r+D_r$) state is absent.
(b) Orientation autocorrelation $\langle\uv(\tau)\cdot\uv(0)\rangle_r$ for $r=0.01,~0.5,$ and $10$ at $\Omega_0=2.5$(marked symbols in (a)); analytic predictions from Eq.~\eqref{eq:ncorr_resetting} are shown as solid lines and simulation as points.
(c) Plot of steady-state radial position distribution $P_{\rm st}(|\rv|)$ for the same parameter values as (b); points are from simulation and dashed lines are corresponding Gaussian($P^{G}_{\rm st}(|\rv|)$) using MSD $\la\rv^2\ra_{r}^{\rm st}$(Eq.~\eqref{eq:r2avg_reset_st}).
(d,e,f) Plot of the steady-state position distribution $P_{\rm st}(x,y)$.
Fixed parameters are $v_0=20$, $D_r=0.05$, and $D=1$.
} 
\label{fig4}
\end{center}
\end{figure*}

\medskip
\section{state diagram}
\label{sec:state_diagrams}
For a Gaussian process with zero mean $\la\rv\ra_{r}=0$, fourth moment
obeys the relation $\mu^{r}_{4}=2\la\rv^2\ra_{r}^2$ with steady state $(\mu^{r}_{4})_{\rm st}=2(\la\rv^2\ra^{\rm st}_{r})^2$(see Eq.~\eqref{eq:r2avg_reset_st} for $\la\rv^2\ra^{\rm st}_{r}$).
We introduce the excess kurtosis $\mathcal{K}_{r}=(\la\rv^4\ra_{r}/\mu^{r}_{4}) - 1$, where the calculation of $\la\rv^4\ra_{r}$ to quantify deviations from Gaussian-like behavior, and thus measure the degree of active motion in the steady-state distribution, shown explicitly in Appendix~\ref{app:fourth_moment}.
At long times, We calculate steady-state excess kurtosis $\mathcal{K}^{\rm st}_{r} = \mathcal{K}_{r} (t\to\infty)$, 
\bea
\mathcal{K}^{\rm st}_{r}=\f{\la\rv^4\ra^{\rm st}_{r}}{(\mu^{r}_{4})_{\rm st}} - 1\,,
\label{eq:excess_kurtosis_st}
\eea
where $(\mu^{r}_{4})_{\rm st}=2(\la\rv^2\ra^{\rm st}_{r})^2$ with $\la\rv^2\ra^{\rm st}_{r}$ presented in Eq.~\eqref{eq:r2avg_reset_st} and $\la\rv^4\ra^{\rm st}_{r}$ calculated in Appendix~\ref{app:fourth_moment} reads
\begin{widetext}
\bea
\la\rv^4\ra_{r}^{\rm st} &=& \frac{64 D^2}{r^{2}} + \frac{32 D N_2 v_0^2}{r^{2} \big[(r+D_r)^2 + \W_0^2\big]^2}+ \frac{8 N_4 v_0^4}{r^{2} \big[(r+D_r)^2 + \W_0^2\big]^2 \big[4 \W_0^2 + (4 D_r + r)^2\big]}\,,\nonumber\\
N_2 &=& 2 D_r^3 + 2 D_r \W_0^2 + 7 r D_r^2  + r \W_0^2  + 8 r^2 D_r  + 3 r^3\,,\nonumber\\
N_4 &=& 32 D_r^4 + 84 r D_r^3 + 75 r^2 D_r^2 + 26 r^3 D_r + 3 r^4 + 8 D_r^2 \W_0^2 + 8 r D_r \W_0^2 + 3 r^2 \W_0^2\,.
\label{eq:r4avg_reset_st}
\eea
\end{widetext}
We use Eq.~\eqref{eq:excess_kurtosis_st} to classify the steady-state position distribution as light-tailed($\mathcal{K}^{\rm st}_{r}<0$) when active motion dominates and heavy tailed($\mathcal{K}^{\rm st}_{r}>0$) when resetting dominates.
%
The combination of spatial($\mathcal{K}^{\rm st}_{r}$) and temporal($\la\uv(\tau)\cdot\uv(0)\ra_r$) diagnostics gives a clear, intuitive picture of how the particle explores space under competing influences.  
The excess kurtosis tells us whether long excursions between resets (heavy-tailed) or smooth circular wandering (light-tailed) dominate the steady-state displacement distribution.  
Meanwhile the orientation autocorrelation $\la\uv(\tau)\cdot\uv(0)\ra_r$ in Eq.~\eqref{eq:ncorr_resetting} pinpoints whether the particle retains its heading long enough to complete many loops (oscillatory decay) or forgets its direction almost immediately (monotonic decay).

These two measures together map out a state diagram(Fig.~\ref{fig4}(a)) in which the relative rates of resetting, rotation, and diffusion cleanly delineate three distinct spatiotemporal behaviors.

\begin{itemize}
\item \textbf{Resetting I:} Occasional resets interrupt scattered loops without fully suppressing them(see trajectory in Fig.~\ref{fig1}(a)).  The distribution develops heavy tails from rare long excursions with autocorrelation exhibits damped oscillations.

\item \textbf{Active:} Self-propulsion and chirality dominate.  The particle follows persistent circular arcs with concentrated loop(see trajectory in Fig.~\ref{fig1}(b)), the displacement distribution remains narrow (light-tailed), and the orientation autocorrelation exhibits damped oscillations.

\item \textbf{Resetting II:} Frequent resets (or weak chirality) effectively quench circular motion(see trajectory in Fig.~\ref{fig1}(c)).  The particle executes short, nearly straight flights between resets, producing a heavy-tailed distribution, and the autocorrelation decays monotonically.
\end{itemize}

\medskip

The achiral ABP exhibit only the `Resetting II' state, with excess kurtosis varying between $1 < \mathcal{K}^{\mathrm{st}}_{r} < 2$ and a re-entrant behavior in $r$, as reported in Shee {\em et al.}~\cite{Shee2025}. In contrast, chirality suppresses the excess kurtosis and leads to three distinct states: the `Resetting II' state, identical to the achiral case~\cite{Shee2025}, and two additional states — `Resetting I' with $0 < \mathcal{K}^{\mathrm{st}}_{r} < 1$, and an `Active' state with $-0.25 < \mathcal{K}^{\mathrm{st}}_{r} < 0$ characterized by chiral rotation. Thus, chirality reduces the effective exploration of space compared to the achiral (or non-chiral) counterpart in active systems under stochastic resetting.
Furthermore, compared with chiral ABP in a harmonic trap~\cite{Pattanayak2025}, which display only weakly heavy-tailed distributions with chiral rotation, our resetting model exhibits strongly heavy-tailed regimes, both with and without chiral rotation. This demonstrates that stochastic resetting significantly amplifies tail behavior while also suppressing chiral rotation.

\medskip
In the absence of both translational and rotational noise ($D = D_{r} = 0$), Eq.~\eqref{eq:r4avg_reset_st} simplifies to,
\bea
\la\rv^4\ra_{r}^{\rm st} &=& \frac{24 v_0^4}{(r^2 + \W_0^2) (4 \W_0^2 + r^2)}\,.
\label{eq:r4avg_reset_st_noiseless_limit}
\eea
From this one finds that the steady-state excess kurtosis(using Eq.~\eqref{eq:excess_kurtosis_st}) becomes
$\mathcal{K}^{\rm st}_{r}=(2 r^2 - \W_0^2)/(4 \W_0^2 + r^2)$.
In this limit, the excess kurtosis attains a minimum of $-1/4$ (weakly light-tailed) as $r\to0$, and a maximum of $2$ (heavy-tailed) as $\Omega_0\to0$.
Thus even in the idealized noiseless setting, the competition between deterministic chirality and stochastic resetting alone can tune the distribution from light- to heavy-tailed. 
Table~\ref{tab:regime_limits} summarizes whether each of the three regimes appears or vanishes under the various limiting cases with additional figures in Appendix.

\medskip
Figure~\ref{fig3} compares the analytic prediction for the steady‑state excess kurtosis (solid lines, Eq.~\eqref{eq:excess_kurtosis_st}) with numerical results (symbols), showing excellent agreement.
At low rotational diffusion coefficients the system enters a weakly active regime (negative excess kurtosis).
This regime arises from the interplay of chirality and stochastic resetting and is not observed for achiral active Brownian particles under resetting~\cite{Shee2025}.
We identify three distinct regimes based on the steady‐state excess kurtosis $\mathcal{K}^{\mathrm{st}}_{r}$: a weakly active state when $\mathcal{K}^{\mathrm{st}}_{r}<0$, Resetting I (weak resetting) for $0<\mathcal{K}^{\mathrm{st}}_{r}<1$, and Resetting II (strong resetting) when $\mathcal{K}^{\mathrm{st}}_{r}>1$.
The Resetting II regime mirrors the achiral case~\cite{Shee2025}, exhibiting a re‑entrant dependence on the resetting rate.

\begin{table*}[!t]
\centering
\caption{Presence (\checkmark) or absence (\texttimes) of the three regimes in various limiting cases. Here, `Re-entrant' indicates the transition from `Resetting I' to `Active' to `Resetting II' state.}
\begin{tabular}{lcccc}
\hline
Limit & Active & Resetting I & Resetting II & Remarks \\
\hline
Our model  & \checkmark & \checkmark & \checkmark & $-0.25<\mathcal{K}^{\mathrm{st}}_{r}<2$; Re-entrant in both $\Omega_0$ and $r$ \\
No translational noise ($D=0$)                  & \checkmark & \checkmark & \checkmark & $-0.25<\mathcal{K}^{\mathrm{st}}_{r}<2$; Re-entrant in $r$ only \\
Noiseless ($D=0,\;D_r=0$)                    & \checkmark & \texttimes & \checkmark &  $-0.25<\mathcal{K}^{\mathrm{st}}_{r}<2$; No Re-entrance\\
Achiral(or non-chiral) ($\Omega_0=0$)~\cite{Shee2025}                 & \texttimes & \texttimes & \checkmark &  $1<\mathcal{K}^{\mathrm{st}}_{r}<2$; Re-entrant in $r$\\
Brownian ($\Omega_0=0,~v_0=0$)~\cite{Shee2025}                 & \texttimes & \texttimes & \checkmark &  $\mathcal{K}^{\mathrm{st}}_{r}=1$; constant\\
Free CABP ($r=0$)~\cite{Liebchen2022, Pattanayak2024}                    & \texttimes & \texttimes & \texttimes & No steady state  \\
Free ABP ($r=0$, $\W_0=0$)~\cite{Basu2018, Shee2020}                    & \texttimes & \texttimes & \texttimes & No steady state \\
\hline
\end{tabular}
\label{tab:regime_limits}
\end{table*}

\medskip
The spatiotemporal state diagram in the $r–\Omega_0$ plane is shown in Fig.~\ref{fig4}(a), with the black solid line $r + D_r = \Omega_0$ marking the boundary between oscillatory temporal behavior for $r < \Omega_0 - D_r$ and non‑oscillatory behavior for $r > \Omega_0 - D_r$.
The color map in Fig.~\ref{fig4}(a) represents the steady‑state excess kurtosis $\mathcal{K}^{\mathrm{st}}_{r}$; the red dashed line indicates $\mathcal{K}^{\mathrm{st}}_{r}=0$(Gaussian), and the green dash‑dotted line indicates $\mathcal{K}^{\mathrm{st}}_{r}=1$(heavy-tailed correspond to Brownian particle under stochastic resetting; see Table~\ref{tab:regime_limits}). {\em Note}, $\mathcal{K}^{\mathrm{st}}_{r}=0$ corresponds to a Gaussian-like position distribution arising from circular trajectories($r + D_r < \Omega_0$); therefore, this is not a passive state.
The close alignment of the lines $r + D_r = \Omega_0$ and $\mathcal{K}^{\mathrm{st}}_{r}=1$ indicates that strong resetting (Resetting II) emerges only when chiral rotation is suppressed and the particle behaves effectively as an achiral particle~\cite{Shee2025}.
In the weak resetting regime (Resetting I), the particle exhibits a heavy‑tailed position distribution while undergoing chiral rotation, whereas in the weakly active state it displays a weak light‑tailed distribution, also with chiral rotation.
These temporal and spatial distinctions at steady state are further illustrated by the orientation autocorrelation curves in Fig.~\ref{fig4}(b), the radial position distributions in Fig.~\ref{fig4}(c), and heat map of position distributions in $x$-$y$ plane in Figs.~\ref{fig4}(d-f)  with trajectories shown in Figs.~\ref{fig1}(b-d).

In Fig.~\ref{fig4}(b) (and in Appendix Fig.~\ref{app_fig1}(a,b)), the solid lines show the predictions of Eq.~\eqref{eq:ncorr_resetting}, which are in excellent agreement with the simulation results (points). It clearly shows oscillatory (chiral) motion at low resetting rates and non-oscillatory behavior at high resetting rates, consistent with the state diagram.
In Fig.~\ref{fig4}(c), we plot the steady-state radial distribution $P_{\rm st}(|\rv|)$ calculated from direct simulation(points). To visualize the non-equilibrium deviation due to the interplay of chiral activity and resetting, we plot the reference Gaussian  $P^{G}_{\rm st}(|\rv|) = e^{-\rv^2/2\sigma^2} / 2\pi \sigma^2$ in dashed lines with effective standard deviation $\sigma=\sqrt{\la\rv^2\ra^{\rm st}_r/2}$. The heavy-tailed distributions at both low and high resetting rates indicate re-entrant behavior, with a weak light-tailed distribution emerging at intermediate resetting rates.

\section{Conclusions}
\label{sec:conclusions}

In this work, we have analyzed the dynamics of a chiral active Brownian particle (CABP) subject to stochastic resetting, in which both its position and orientation are restored to predetermined values at a constant rate.
Using a renewal equation framework together with a Fokker–Planck formalism, we derived analytical expressions for the particle’s orientation dynamics and displacement moments, including the mean squared displacement and the fourth order displacement moment.
Our analysis demonstrates that the interplay between the resetting rate, intrinsic chirality, and rotational diffusion gives rise to spatiotemporal behaviors.
In the presence of chirality, we identify three states: active, resetting I, and resetting II, whereas in an achiral system only the resetting state is observed~\cite{Shee2025}.
Thus, chirality suppresses the heavy tailed position distribution and produces a re-entrant transition from resetting to an active regime and back to resetting when either the resetting rate or chirality is varied independently.

Numerical simulations performed using the Euler-Maruyama method corroborate our theoretical predictions, demonstrating excellent agreement across both transient and steady-state regimes.
Our findings underscore that stochastic resetting acts as a versatile control parameter, enabling the tuning of transport properties and the manipulation of non-equilibrium steady states in active matter systems with rotational degrees of freedom, paving the way for single-particle experimental realization~\cite{MunozGil2025}.

Resetting can be experimentally implemented across several established platforms~\cite{Tal-Friedman2020}. Optical tweezers can reposition or reorient Janus colloids with high precision, magnetic fields can steer and realign magnetically coated chiral swimmers, light pulses can trigger controlled tumbling events in engineered phototactic bacteria, and microfluidic flow-switching architectures can return microswimmers to predefined reset locations~\cite{Tkachenko2023, Duygu2025, Rey2023}. These approaches enable controlled position–and–orientation resets and allow direct testing of our predictions. Moreover, our theoretical predictions can be tested at larger scales using macroscopic robotic active-matter experiments, which effectively operate in the overdamped limit (or can be approximated by neglecting inertial effects) and allow precise control over resetting protocols~\cite{Dauchot2019, Pramanick2024, Olsen2025}.

The insights presented here deepen our understanding of non-equilibrium phenomena in chiral active systems and open new avenues for future investigation. 
It is also important to analyze the transient dynamics in detail (MSD in Eq.~\eqref{eq:r2avg_reset}; Appendix~\ref{app:MSD}, fourth moment in Appendix~\ref{app:fourth_moment}, and excess kurtosis in Appendix~\ref{app:excess_kurtosis} as a function of time) to understand the complex dynamical behavior. However, we have not pursued this here, as our paper focuses solely on the steady state.
Extending this framework to inertial systems~\cite{PatelChaudhuri2023, PatelChaudhuri2024}, many-particle systems~\cite{NagarGupta2023, VatashRoichman2025}, investigating the effects of external fields and confinement~\cite{Caprini2019, Abdoli2021, Chaudhuri2021, Caprini2023, PatelChaudhuri2024, Shee2025} could further advance the control and optimization of active transport processes.

\begin{acknowledgments}
We thank Priyo Shankar Pal for valuable discussions during the early stages of this work.
\end{acknowledgments}

\section*{Data availability} 
All data supporting the findings of this study are available within the article.

\bibliography{reference} 

\onecolumngrid
\appendix

\section{Analytic calculations of lower order moments: orientation autocorrelation, mean displacement, position-orientation cross correlations}
\label{app:lower_order_moments}

\subsection{Orientation autocorrelation}
We assume that the CABP is initially oriented along $\uv_0$ (i.e.\ $\langle \uv(0)\rangle = \uv_0$) and proceed to calculate the mean orientation.
To calculate $\la\uv\ra$, we substitute $\psi =\uv$ in the Eq.~\eqref{eq:moment}, leads to Laplace space mean orientation
\bea
\la\uv\ra_s &=& \f{\uv_0 + \W_0 \la\uv^{\perp}\ra_s}{s+D_r}\,,
\eea
where we can calculate $\la\uv^{\perp}\ra_s$ with initial perpendicular orientation $\la\uv^{\perp}(0)\ra=\uv^{\perp}_0$ using $\psi=\uv^{\perp}$ in the Eq.~\eqref{eq:moment} gives
\bea
\la\uv^{\perp}\ra_s &=& \f{\uv^{\perp}_0 - \W_0 \la\uv\ra_s}{s+D_r}\,,
\eea
substituting back into the $\la\uv\ra_s$
\bea
\la\uv\ra_s &=& \f{\left[ \uv_0(s+D_r)+ \W_0 \uv^{\perp}_0\right]}{[(s+D_r)^2+\W_0^2]}\,,
\label{eq:nav_Laplace_active}
\eea
Inverse Laplace transform leads to the average orientation without stochastic resetting
\bea
\la\uv(t)\ra &=& e^{-D_r t} [\uv_0 \cos(\W_0 t)+ \uv^{\perp}_0 \sin(\W_0 t)]\,.
\eea
In the presence of stochastic resetting, using Eq.~\eqref{eq:moment_resetting}, we get
\bea
\la\uv(t)\ra_r &=& e^{-(r+D_r) t} [\uv_0 \cos(\W_0 t)+ \uv^{\perp}_0 \sin(\W_0 t)]\nonumber\\
&& + r\uv_0 \f{(r+D_r)[1-e^{-(r+D_r)t}\cos(\W_0 t)]+\W_0 e^{-(r+D_r)t} \sin(\W_0 t)}{(r+D_r)^2+\W_0^2}\nonumber\\
&& + r\uv^{\perp}_0 \f{\W_0[1-e^{-(r+D_r)t}\cos(\W_0 t)]- (r+D_r) e^{-(r+D_r)t} \sin(\W_0 t)}{(r+D_r)^2+\W_0^2}\,.
\eea
Taking the dot product with the initial orientation yields the steady state orientation autocorrelation $\la\uv(\tau)\cdot\uv(0)\ra_r\equiv\la\uv(t)\cdot\uv(0)\ra_r$ given in Eq.~\eqref{eq:ncorr_resetting} of the main text(see also Fig.~\ref{fig4}(b)).
At long timescale($\tau\to\infty$), the steady state orientation autocorrelation saturates to finite value $C=\la\uv(\tau)\cdot\uv(0)\ra_r (\tau\to \infty)$ as reported in Eq.~\eqref{eq:ucorr_st} of the main text. Equivalently, $C$ can be obtained directly by applying the Final Value Theorem to Eq.~\eqref{eq:nav_Laplace_active} via Eq.~\eqref{eq:moment_resetting_FVT}.

\begin{figure}[!t]
\begin{center}
\includegraphics[width=0.5\linewidth]{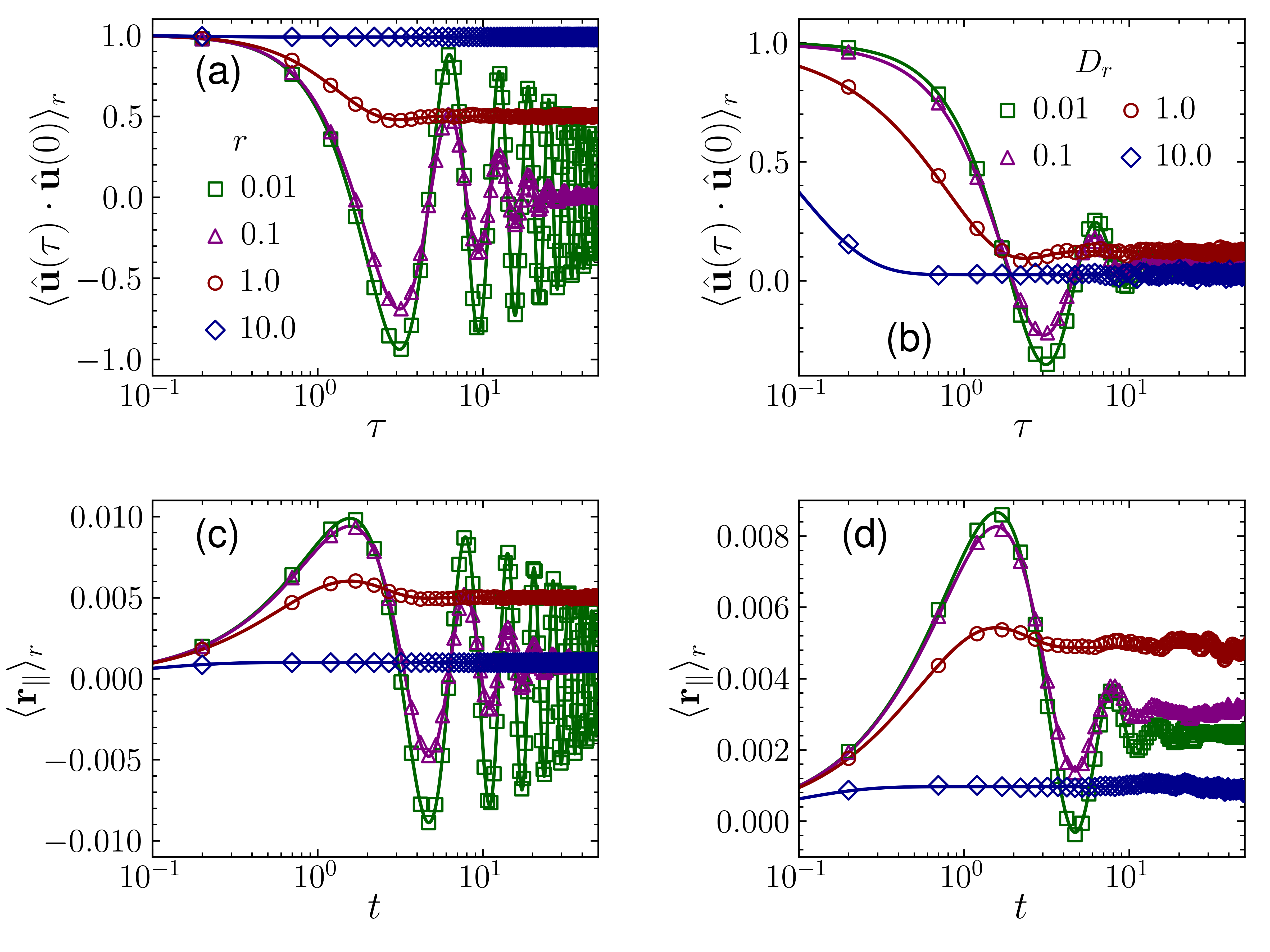} 
\caption{
(a,b) Orientation autocorrelation $\la\uv(\tau)\cdot\uv(0)\ra_{r}$ and (c,d) mean parallel displacement $\la\rv_{\parallel}\ra_{r}$ along initial orientation direction in two dimension under stochastic resetting. The points are from simulation. The solid lines are the plot of Eq.~\eqref{eq:ncorr_resetting} in (a,b) and Eq.~\eqref{eq:rpara_resetting} in (c,d). The initial position is at the origin with the initial orientation along the $x$-axis. Fixed parameter are $v_0=0.01$, $\W_0=1$, $D=1$ with $D_r=0.01$ in (a,c) and $r=0.25$ in (b,d).
} 
\label{app_fig1}
\end{center}
\end{figure}

In Fig.~\ref{app_fig1}(a,b), the solid lines represent the predictions of Eq.~\eqref{eq:ncorr_resetting}, which are in excellent agreement with the simulation results (points).
We observe that increasing either the resetting rate or the rotational diffusion coefficient suppresses the oscillatory behavior of the orientation autocorrelation.
However, increasing the resetting rate pushes the long‑time saturation level to unity, whereas increasing $D_r$ first raises (maximum at $D_r^{*}=\W_0-r$ if $\W_0>r$) and then lowers the steady‑state autocorrelation.

\subsection{Mean displacement}
To calculate $\la\rv\ra_s$, we set the initial position at $\la\rv\ra_0=0$ and substitute $\psi=\rv$ into Eq.~\eqref{eq:moment} leads to
\bea
\la\rv\ra_s &=& \f{v_0 \la\uv\ra_s}{s}\,.
\label{eq:rav_Laplace}
\eea
Substituting the expression for $\la\uv\ra_s$ from Eq.~\eqref{eq:nav_Laplace_active} into Eq.~\eqref{eq:rav_Laplace} and performing the inverse Laplace transform gives
\bea
\la\rv(t)\ra &=& \f{v_0 \uv_0}{D_r^2+\W_0^2} \left[D_r[1-e^{-D_r t}\cos(\W_0 t)]+\W_0 e^{-D_r t} \sin(\W_0 t)\right]\nn\\
& &+ \f{v_0 \uv^{\perp}_0}{D_r^2+\W_0^2} \left[\W_0 [1- e^{-D_r t}\cos(\W_0 t)]-D_r e^{-D_r t} \sin(\W_0 t)\right]\,.
\eea
We decompose the displacement vector $\rv$ into components parallel and perpendicular to the initial orientation $\uv_0$ as
\bea
\rv_{\parallel} &= (\rv\cdot\uv_0)\uv_0~~,~~\rv_{\perp} &=\rv - \rv_{\parallel}\,.
\eea
The component of the displacement vector parallel to the initial orientation is given by
\bea
\la\rv_{\parallel}(t)\ra &=& \f{v_0 \uv_0}{D_r^2+\W_0^2} \left[D_r[1-e^{-D_r t}\cos(\W_0 t)]+\W_0 e^{-D_r t} \sin(\W_0 t)\right]\,.
\label{eq:rpara}
\eea  
With stochastic resetting (Eq.~\eqref{eq:moment_resetting}), one obtains
\bea
\la\rv_{\parallel}(t)\ra_r &=& e^{-r t} \la\rv_{\parallel}\ra + \f{v_0 D_r(1-e^{-rt})\uv_0}{(D_r^2+\W_0^2)}\nonumber\\
&& - \f{rv_0 D_r\uv_0}{(D_r^2+\W_0^2)}\f{(r+D_r)[1-e^{-(r+D_r)t}\cos(\W_0 t)]+\W_0 e^{-(r+D_r)t} \sin(\W_0 t)}{(r+D_r)^2+\W_0^2}\nonumber\\
&& + \f{rv_0 \W_0\uv_0}{(D_r^2+\W_0^2)}\f{\W_0[1-e^{-(r+D_r)t}\cos(\W_0 t)] - (r+D_r) e^{-(r+D_r)t} \sin(\W_0 t)}{(r+D_r)^2+\W_0^2}\,.
\label{eq:rpara_resetting}
\eea  
In the long‑time limit  $t\to\infty$, the steady‑state mean parallel displacement is $\la\rv_{\parallel}\ra^{\rm st}_r = \la\rv_{\parallel}(t)\ra_r(t\to\infty)$, reads
\bea
\la\rv_{\parallel}\ra^{\rm st}_r &=& \f{v_0(r+D_r)}{(r+D_r)^2+\W_0^2}\,.
\label{eq:rpara_st}
\eea  
Equivalently, $\la\rv_{\parallel}\ra^{\rm st}_r$ can be obtained directly by applying the Final Value Theorem to Eq.~\eqref{eq:rav_Laplace} via Eq.~\eqref{eq:moment_resetting_FVT}.

\subsection{Parallel position-orientation cross-correlation}
To compute the position–orientation cross‐correlation $\la\rv\cdot\uv\ra$, We impose the initial conditions $\la\rv\ra=\rv_0$ and $\uv=\uv_0$, and substitute $\psi=\rv\cdot\uv$ into Eq.~\eqref{eq:moment}, leads to
\bea
\la\rv\cdot\uv\ra_s &=& \f{ v_0 \la 1\ra_s + \W_0 \la\rv\cdot\uv^{\perp}\ra_s}{s+D_r}\,,
\label{eq:rn_Laplace1}
\eea
where $\la 1\ra_s=\int_{0}^{\infty} e^{-st} dt=1/s$. In similar method, we compute $\la\rv\cdot\uv^{\perp}\ra_s$ using Eq.~\eqref{eq:moment}, leads to
\bea
\la\rv\cdot\uv^{\perp}\ra_s &=& \f{ - \W_0 \la\rv\cdot\uv\ra_s}{s+D_r}\,.
\label{eq:rnperp_Laplace1}
\eea
Finally, by substituting Eq.~\eqref{eq:rnperp_Laplace1} into Eq.~\eqref{eq:rn_Laplace1}, we obtain the Laplace form of $\la\rv\cdot\uv\ra_s$,
\bea
\la\rv\cdot\uv\ra_s &=& \f{(s+D_r) v_0 \la 1\ra_s}{(s+D_r)^2+\W_0^2}\,.
\label{eq:rn-Laplace}
\eea
The inverse Laplace transform yields the following equation
\bea
\la\rv\cdot\uv\ra(t) &=& \f{v_0}{D_r^2+\W_0^2}\left[D_r\left(1-e^{-D_r t}\cos(\W_0 t)\right)+\W_0 e^{-D_r t}\sin(\W_0 t)\right].
\label{eq:rnavg_time}
\eea
In the presence of stochastic resetting, using Eq.~\eqref{eq:moment_resetting}, we get 
\bea
\la\rv\cdot\uv\ra_{r}(t) &=& e^{-r t} \la\rv\cdot\uv\ra(t) + \f{v_0 D_r(1-e^{-rt})}{(D_r^2+\W_0^2)}\nonumber\\
&& - \f{rv_0 D_r}{(D_r^2+\W_0^2)}\f{(r+D_r)[1-e^{-(r+D_r)t}\cos(\W_0 t)]+\W_0 e^{-(r+D_r)t} \sin(\W_0 t)}{(r+D_r)^2+\W_0^2}\nonumber\\
&& + \f{rv_0 \W_0}{(D_r^2+\W_0^2)}\f{\W_0[1-e^{-(r+D_r)t}\cos(\W_0 t)] - (r+D_r) e^{-(r+D_r)t} \sin(\W_0 t)}{(r+D_r)^2+\W_0^2}\,.
\label{eq:cross-corr}
\eea  
The position–orientation cross correlation $\langle \mathbf r\cdot\mathbf u\rangle_r$ coincides with the mean displacement component parallel to the initial orientation, as given by Eq.~\eqref{eq:rpara_resetting}.
At long times $t\to\infty$, steady state position-orientation cross-correlation $\la\rv\cdot\uv\ra_{r}^{\rm st} = \la\rv\cdot\uv\ra_{r} (t\to\infty)=\la\rv_{\parallel}\ra^{\rm st}_r\cdot\uv_0$(see Eq.~\eqref{eq:rpara_st}).
Equivalently, $\la\rv\cdot\uv\ra_{r}^{\rm st}$ can be obtained directly by applying the Final Value Theorem to Eq.~\eqref{eq:rn-Laplace} via Eq.~\eqref{eq:moment_resetting_FVT}.

\subsection{Perpendicular position–orientation cross-correlation}

We obtain the perpendicular position–orientation cross-correlation by substituting Eq.~\eqref{eq:rn_Laplace1} into Eq.~\eqref{eq:rnperp_Laplace1} in Laplace space,
\bea
\la\rv\cdot\uv^{\perp}\ra_s &=& \f{-\W_0 v_0\la 1\ra_s}{(s+D_r)^2+\W_0^2}\,.
\label{eq:rnperp-Laplace}
\eea
where $\la 1\ra_s=1/s$.
Applying an inverse Laplace transform results in
\bea
\la\rv\cdot\uv^{\perp}\ra (t) &=&  -\f{v_0}{D_r^2+\W_0^2}\left[\W_0 \left(1-e^{-D_r t}\cos(\W_0 t)\right)- D_r e^{-D_r t}\sin(\W_0 t)\right]\,.
\label{eq:rnperpavg_time}
\eea
With stochastic resetting
\bea
\la\rv\cdot\uv^{\perp}\ra_{r}(t) &=& e^{-r t} \la\rv\cdot\uv^{\perp}\ra(t) - \f{v_0 \W_0(1-e^{-rt})}{(D_r^2+\W_0^2)}\nonumber\\
&& + \f{rv_0 \W_0}{(D_r^2+\W_0^2)}\f{(r+D_r)[1-e^{-(r+D_r)t}\cos(\W_0 t)]+\W_0 e^{-(r+D_r)t} \sin(\W_0 t)}{(r+D_r)^2+\W_0^2}\nonumber\\
&& + \f{rv_0 D_r}{(D_r^2+\W_0^2)}\f{\W_0[1-e^{-(r+D_r)t}\cos(\W_0 t)] - (r+D_r) e^{-(r+D_r)t} \sin(\W_0 t)}{(r+D_r)^2+\W_0^2}\,.
\label{eq:mean_disp_perp}
\eea  
In the long‑time limit $t\to\infty$, the steady‑state cross correlation $\la\rv\cdot\uv^{\perp}\ra_{r}^{\rm st} = \la\rv\cdot\uv^{\perp}\ra_{r} (t\to\infty)$
\bea
\la\rv\cdot\uv^{\perp}\ra_{r}^{\rm st} &=& -\f{v_0\W_0}{(r+D_r)^2+\W_0^2}\,.
\label{eq:rperp_st}
\eea 
Equivalently, $\langle \mathbf r\cdot\mathbf u^\perp\rangle_r^{\mathrm{st}}$ can be obtained directly by applying the Final Value Theorem to Eq.~\eqref{eq:rnperp-Laplace} via Eq.~\eqref{eq:moment_resetting_FVT}.
The steady‑state cross‑correlation $\langle \mathbf r\cdot\mathbf u^\perp\rangle_r^{\mathrm{st}}$ characterizes the direction of chiral rotation, as it is odd under the transformation $\Omega_0\!\leftrightarrow\!-\Omega_0$.

\section{Calculation of mean squared displacement (MSD)}
\label{app:MSD}

We compute mean-squared displacement (MSD) defining observable $\psi=\rv^2$, utilizing Eq.~\eqref{eq:moment} with initial position at $\rv_0=(0,0)$ leads to MSD in Laplace space $\la\rv^2\ra_s$ 
\bea
\la\rv^2\ra_s &=& \f{1}{s}\left[4 D\la 1\ra_s + 2 v_0 \la\rv\cdot\uv\ra_s\right]\,,
\label{eq:r2avg-Laplace}
\eea
where $\la 1\ra_s=1/s$ and $\la\rv\cdot\uv\ra_s$ already computed in Eq.~\eqref{eq:rn-Laplace}. 
Inverse Laplace transform of Eq.~\eqref{eq:r2avg-Laplace} gives MSD as a function of time $t$
\bea
\la\rv^2\ra(t) &=& 2\left[2D + \f{D_rv_0^2}{D_r^2+\W_0^2}\right]t -\f{2(D_r^2-\W_0^2) v_0^2}{(D_r^2+\W_0^2)^2} + \f{2 v_0^2e^{- D_r t}[(D_r^2-\W_0^2)\cos(\W_0 t)-2D_r\W_0 \sin(\W_0 t)]}{ (D_r^2+\W_0^2)^2}\,.
\label{eq:r2avg}
\eea
In the presence of stochastic resetting, using Eq.~\eqref{eq:moment_resetting}, we get 
\bea
\la\rv^2\ra_r(t) &=& e^{-r t} \la\rv^2\ra(t) + \f{4D}{r} \left(1-(1+rt)e^{-rt}\right)\nonumber\\
&&+\f{2 r v_0^2 }{(D_r^2+\W_0^2)^2} \left[\f{D_r^3 - r D_r^2 + (r+D_r) \W_0^2}{r^2} + \f{(r+D_r)D_r^2-(r+3D_r)\W_0^2}{(r+D_r)^2+\W_0^2}\right]\nonumber\\
&& -\f{2 r v_0^2 e^{-rt}}{r^2(D_r^2+\W_0^2)^2} \left[(1+rt)D_r^3 -r D_r^2 + r \W_0^2 + (1+rt) D_r \W_0^2\right]\nonumber\\
&&-\f{2 r v_0^2 e^{-(r+D_r)t}}{(D_r^2+\W_0^2)^2 ((r+D_r)^2+\W_0^2)} \left[-((r+D_r)D_r^2-(r+3D_r)\W_0^2)\cos(\W_0 t) + \W_0 (3 D_r^2 +2 r D_r -\W_0^2) \sin(\W_0 t)\right]\,.\nonumber\\
\label{eq:r2avg_reset}
\eea
In the long‑time limit $t\to\infty$, the mean‑squared displacement in Eq.~\eqref{eq:r2avg_reset} converges to its steady‑state value $\la\rv^2\ra_{r}^{\rm st}=\la\rv^2\ra_r(t\to\infty)$ as given by Eq.~\eqref{eq:r2avg_reset_st} in the main text.
Equivalently, $\la\rv^2\ra_{r}^{\rm st}$ can be obtained directly by applying the Final Value Theorem to Eq.~\eqref{eq:r2avg-Laplace} via Eq.~\eqref{eq:moment_resetting_FVT}.
In Fig.~\ref{fig2}, we compare the analytic prediction of the steady‑state mean‑squared displacement from Eq.~\eqref{eq:r2avg_reset_st} with the corresponding simulation results.  
A noteworthy observation is that the steady‑state mean‑squared displacement in Eq.~\eqref{eq:r2avg_reset_st} is invariant under chirality reversal $\Omega_0\!\leftrightarrow\!-\Omega_0$.

\begin{figure}[!t]
\begin{center}
\includegraphics[width=0.5\linewidth]{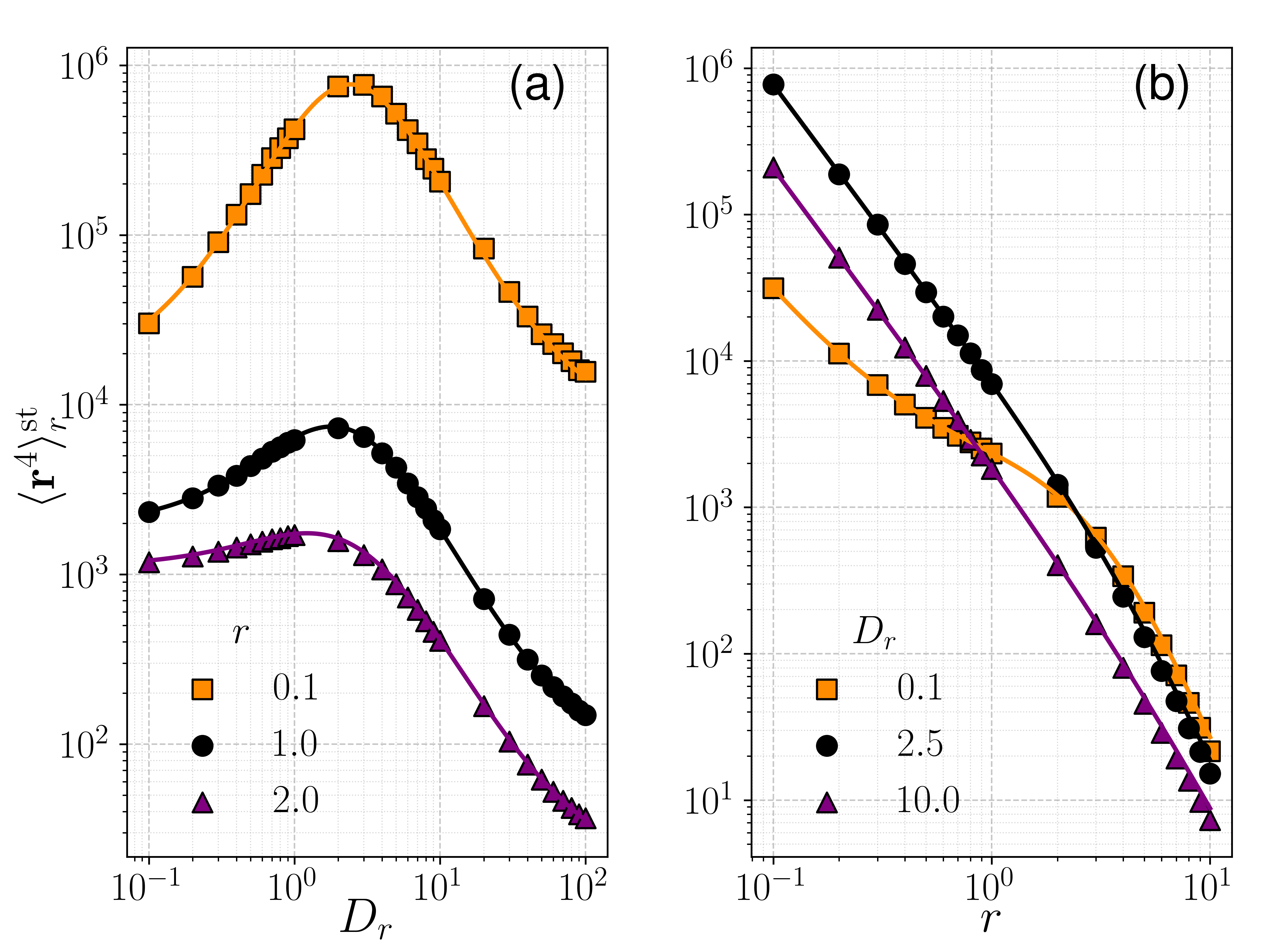} 
\caption{
Steady-state fourth moment of displacement $\la\rv^4\ra_{r}^{\rm st}$ as function of $D_r$ in (a) and of $r$ in (b).
Symbols correspond to simulation, and lines show the prediction from Eq.~\eqref{eq:r4avg_reset_st}.
Fixed parameters are $v_0=10$, $\W_0=2.5$, and $D=1$.
} 
\label{app_fig2}
\end{center}
\end{figure} 
\section{Calculation of fourth order moment of displacement}
\label{app:fourth_moment}
The fourth order moment of displacement in Laplace space with initial position at origin $\rv_0=(0,0)$, using Eq.~\eqref{eq:moment} results
\bea
\la\rv^4\ra_s &=& \f{1}{s}\left[16D\la\rv^2\ra_s + 4 v_0 \la(\uv\cdot\rv)\rv^2\ra_s\right]\,.
\label{eq:r4avg-Laplace}
\eea
Further, we proceed to calculate $\la(\uv\cdot\rv)\rv^2\ra_s$ (see Appendix~\ref{app:urr2} for detailed derivation), reads
\bea
&&\la(\uv\cdot\rv)\rv^2\ra_s = \f{s+D_r}{(s+D_r)^2 +\W_0^2}\left[8D\la\uv\cdot\rv\ra_s+v_0\la\rv^2\ra_s+2 v_0\la(\uv\cdot\rv)^2\ra_s+\frac{\W_0}{s+D_r} \left[8D\la\uv^{\perp}\cdot\rv\ra_s +2 v_0\la(\uv\cdot\rv)(\uv^{\perp}\cdot\rv)\ra_s\right]\right]\,.\nonumber\\
\label{eq:urr2}
\eea
All the required quantity to calculate $\la(\uv\cdot\rv)\rv^2\ra_s$ in fourth order moment listed here
\bea
\la\rv^2\ra_s &=& \f{1}{s}\left[\f{4 D}{s} + 2 v_0 \la\uv\cdot\rv\ra_s\right]\,,\\
\la(\uv\cdot\rv)^2\ra_s &=& \f{s+4 D_r}{(s+4 D_r)^2+(2\W_0)^2}\left[\f{2 D}{s}+2\la\rv^2\ra_s+2 v_0\la\uv\cdot\rv\ra_s+\frac{2\W_0}{s+4 D_r} \left[\W_0\la\rv^2\ra_s+v_0\la\uv^{\perp}\cdot\rv\ra_s\right]\right]\,,\\
\la\uv\cdot\rv\ra_s &=& \f{v_0(s+D_r)}{s[(s+D_r)^2+\W_0^2]}\,,\\
\la\uv^{\perp}\cdot\rv\ra_s &=& \f{- \W_0 \la\uv\cdot\rv\ra_s}{s+D_r}\,,\\
\la(\uv\cdot\rv)(\uv^{\perp}\cdot\rv)\ra_s &=& \f{1}{s+4D_r}\left[\W_0\la\rv^2\ra_s+v_0\la\uv^{\perp}\cdot\rv\ra_s-2\W_0 \la(\uv\cdot\rv)^2\ra_s\right]\,.
\eea

Inverse Laplace transformation of Eq.~\eqref{eq:r4avg-Laplace} leads to fourth moment of displacement; presented full form in Pattanayak {\em et al.}~\cite{Pattanayak2024}. 
In the presence of stochastic resetting, using Eq.~\eqref{eq:moment_resetting} we get fourth moment of displacement under stochastic resetting $\la \rv^4\ra_r(t)$.

\medskip

In the steady state fourth order moment of displacement $\la\rv^4\ra_{r}^{\rm st}=\la\rv^4\ra_r(t\to\infty)$ presented in main text Eq.~\eqref{eq:r4avg_reset_st}.
Equivalently, $\la\rv^4\ra_{r}^{\rm st}$ can be obtained directly by applying the Final Value Theorem to Eq.~\eqref{eq:r4avg-Laplace} via Eq.~\eqref{eq:moment_resetting_FVT}.
Figure~\ref{app_fig2} compares the analytic prediction from Eq.~\eqref{eq:r4avg_reset_st} (solid lines) with simulation results (symbols), demonstrating excellent agreement.
The behavior of $\la\rv^4\ra_{r}^{\rm st}$ is qualitatively analogous to that of the MSD, exhibiting a non‑monotonic dependence on $D_r$ in Fig.~\ref{app_fig2}(a) and a monotonic decay with $r$ in Fig.~\ref{app_fig2}(b).
To quantitatively distinguish between activity‑ and resetting‑dominated regimes, we now compute the excess kurtosis.

\begin{figure}[!t]
\begin{center}
\includegraphics[width=0.5\linewidth]{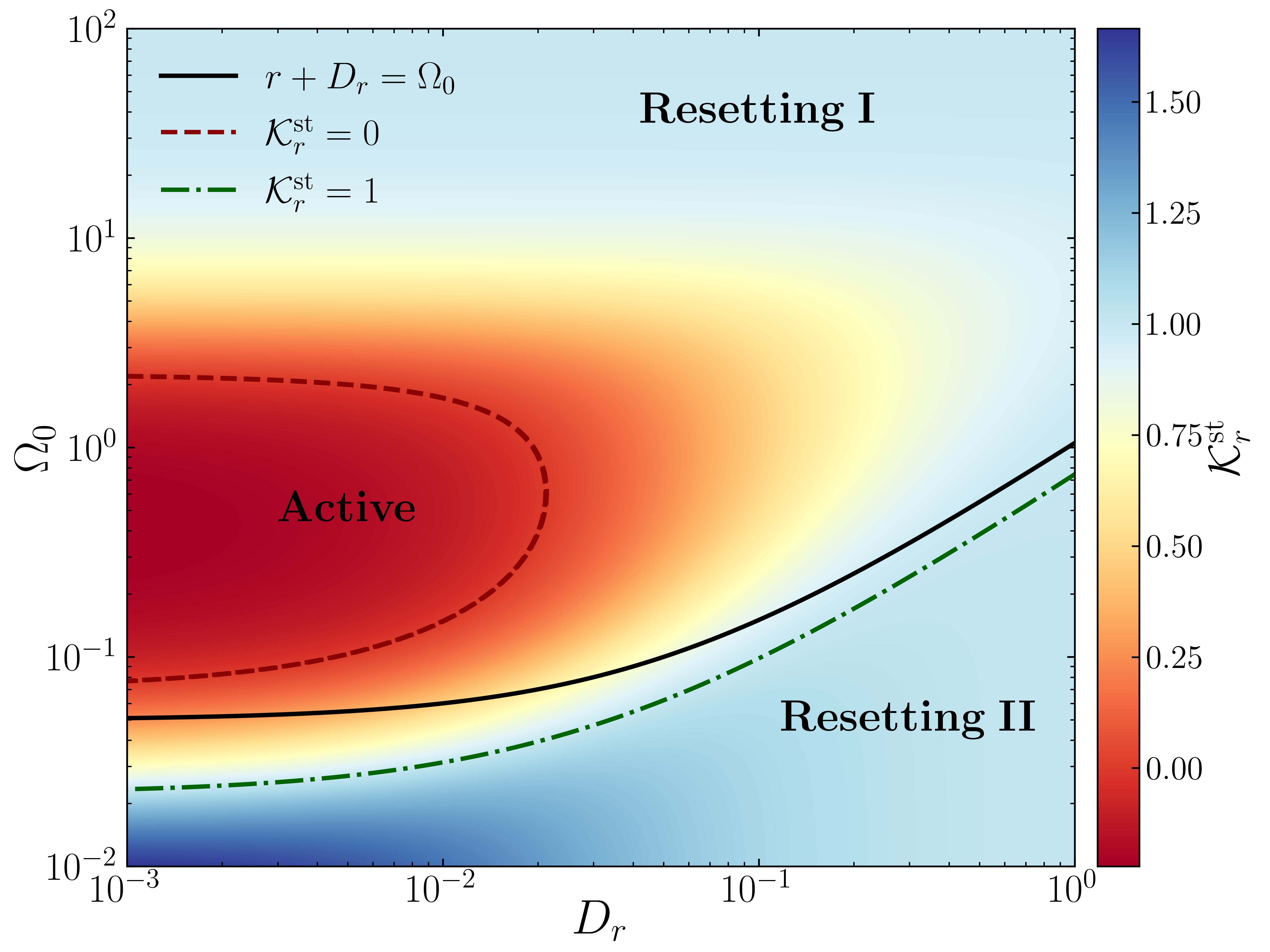} 
\caption{
State diagram in the $D_{r}$–$\Omega_0$ plane.
The colormap depicts the steady-state excess kurtosis $\mathcal{K}_{r}^{\mathrm{st}}$ (Eq.~\eqref{eq:excess_kurtosis_st}).
The black solid line marks the boundary between oscillatory and non-oscillatory orientation dynamics $r+D_r=\W_0$.
The red dashed and green dash-dotted lines represent $\mathcal{K}_{r}^{\mathrm{st}}=0$ and $\mathcal{K}_{r}^{\mathrm{st}}=1$, respectively.
The initial position is at the origin, with the initial orientation along the $x$-axis.
Fixed parameters are $v_0=20$, $r=0.05$, $D=1$.
} 
\label{app_fig3}
\end{center}
\end{figure} 

\begin{figure}[!t]
\begin{center}
\includegraphics[width=0.49\linewidth]{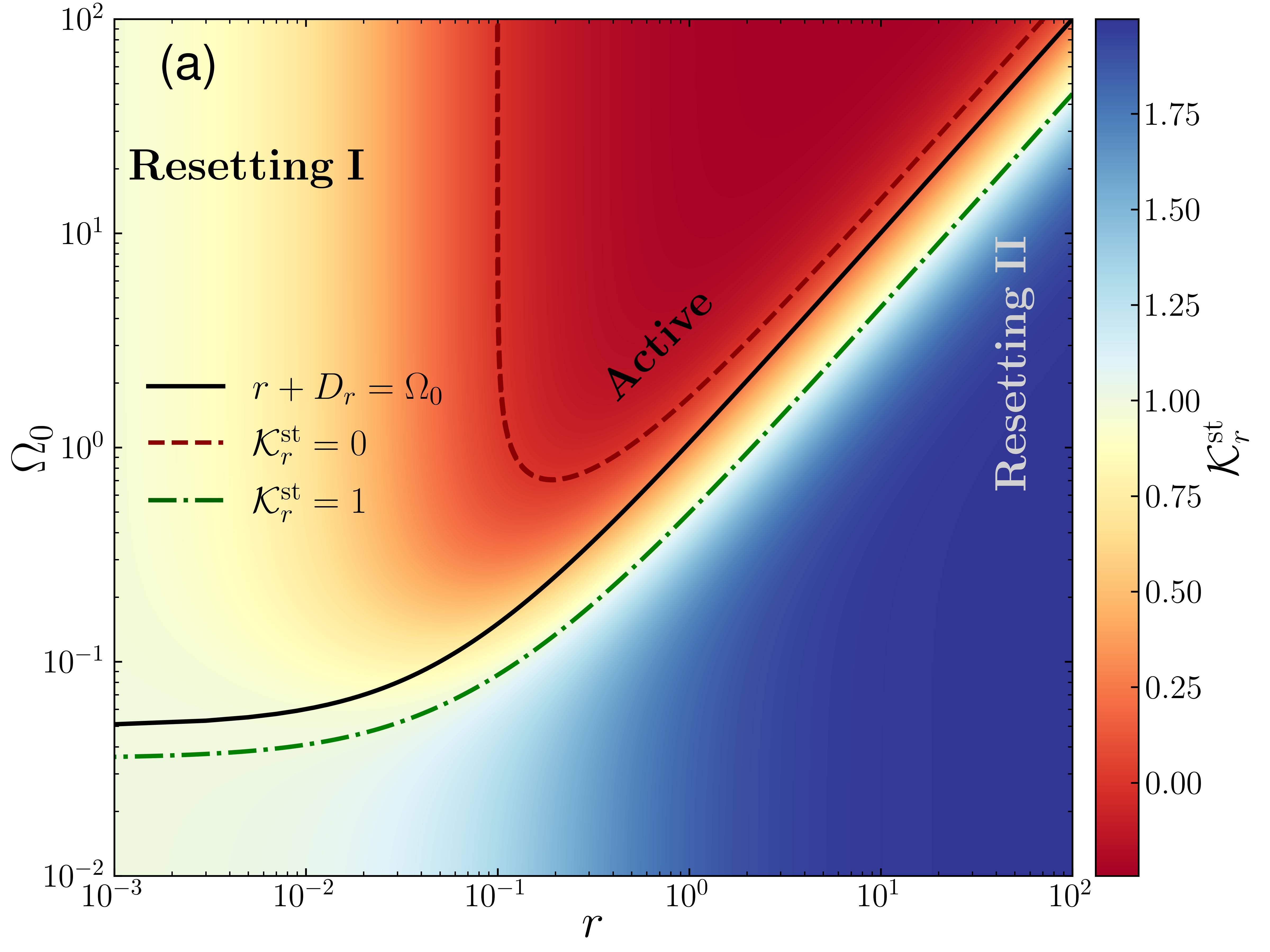} 
\includegraphics[width=0.49\linewidth]{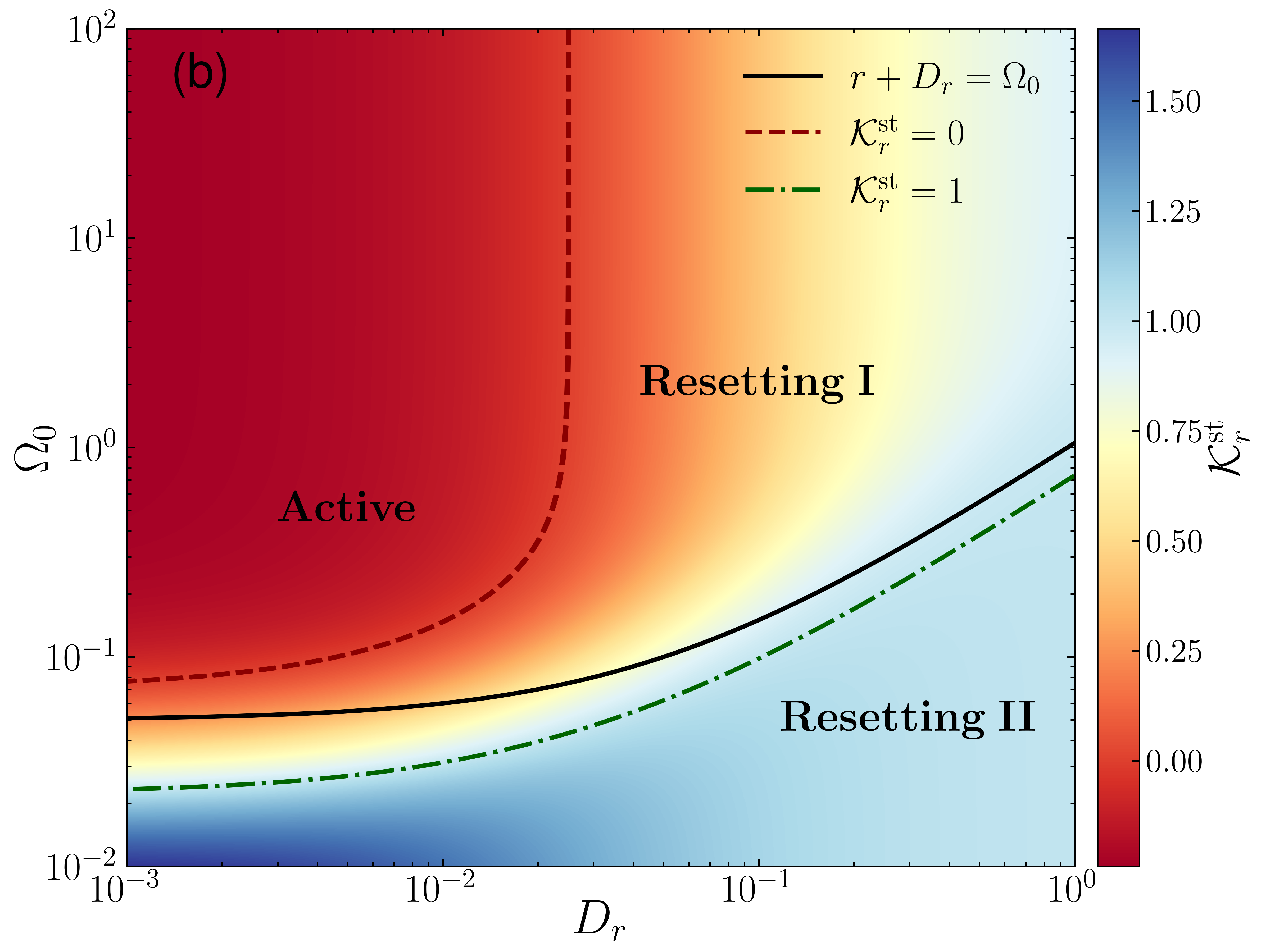} 
\caption{
State diagrams in the (a) $r$–$\Omega_0$ plane and (b) $D_{r}$–$\Omega_0$ plane, in the absence of translational noise ($D=0$).
The colormap depicts the steady-state excess kurtosis $\mathcal{K}_{r}^{\mathrm{st}}$ (Eq.~\eqref{eq:excess_kurtosis_st}).
The black solid line marks the boundary between oscillatory and non-oscillatory orientation dynamics $r+D_r=\W_0$.
The red dashed and green dash-dotted lines represent $\mathcal{K}_{r}^{\mathrm{st}}=0$ and $\mathcal{K}_{r}^{\mathrm{st}}=1$, respectively.
The initial position is at the origin, with the initial orientation along the $x$-axis.
In both panels, $v_0=20$, with $D_{r}=0.05$ in (a) and $r=0.05$ in (b).
} 
\label{app_fig_state_diagrams_D=0}
\end{center}
\end{figure}

\begin{figure}[!t]
\begin{center}
\includegraphics[width=0.5\linewidth]{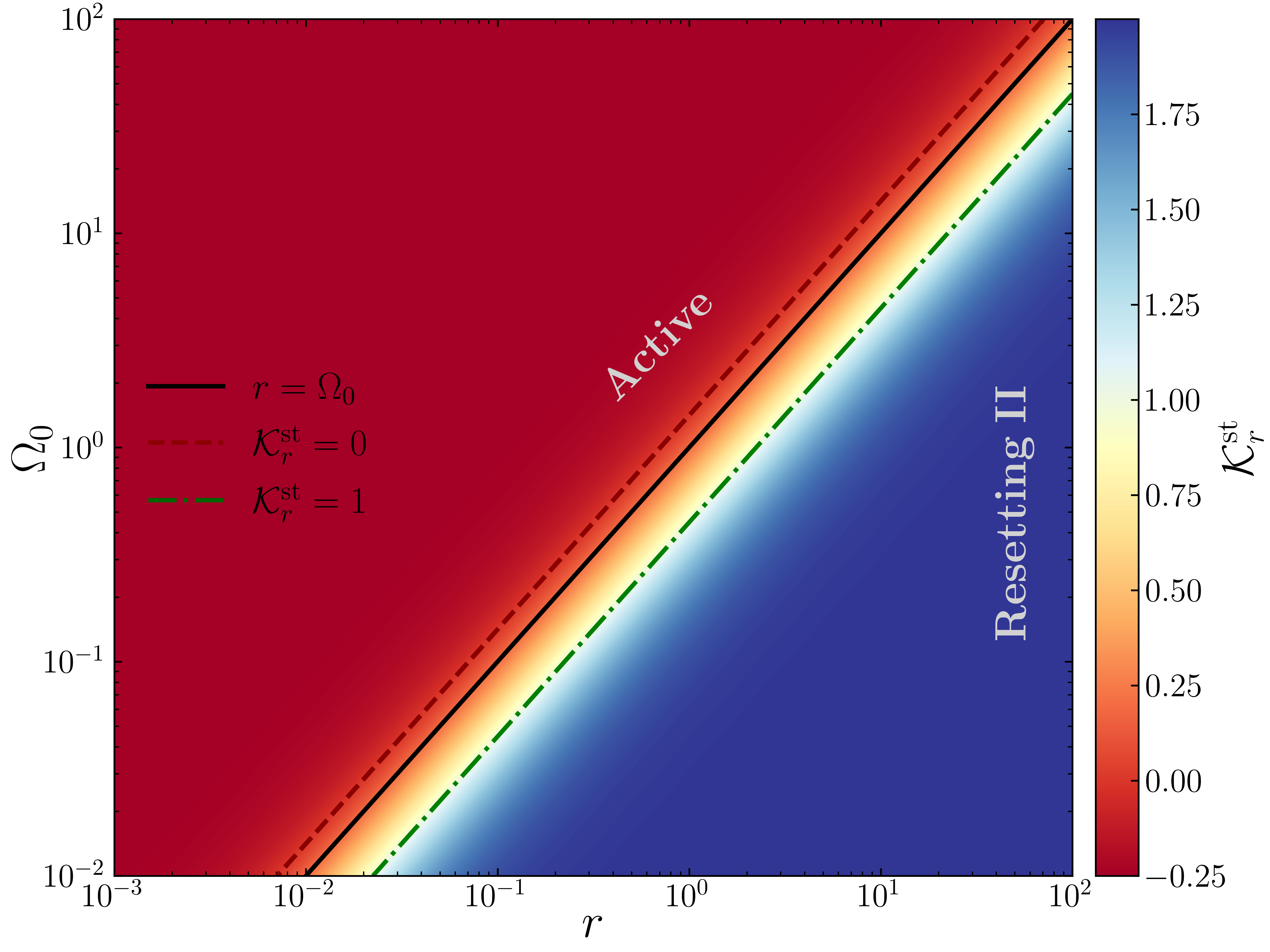} 
\caption{
State diagram in the $r$–$\Omega_0$ plane in the absence of both rotational and translational noise ($D=D_r=0$).
The colormap depicts the steady-state excess kurtosis $\mathcal{K}_{r}^{\mathrm{st}}$ (Eq.~\eqref{eq:excess_kurtosis_st}).
The black solid line marks the boundary between oscillatory and non-oscillatory orientation dynamics $r=\W_0$.
The red dashed and green dash-dotted lines represent $\mathcal{K}_{r}^{\mathrm{st}}=0$ and $\mathcal{K}_{r}^{\mathrm{st}}=1$, respectively.
The initial position is at the origin, with the initial orientation along the $x$-axis.
Fixed parameter is $v_0=20$.
} 
\label{app_fig_state_diagrams_D=0_Dr=0}
\end{center}
\end{figure} 
\section{Calculation of excess kurtosis}
\label{app:excess_kurtosis}
We define kurtosis to capture the deviation of Gaussian nature which signify the how active is our system as follows
\bea
\mathcal{K}_r=\f{\la\rv^4\ra_r}{2(\la\rv^2\ra_r)^2} - 1~,
\eea
where the calculation of $\la\rv^2\ra_r$ and $\la\rv^4\ra_r$ already shown explicitly in Appendix~\ref{app:MSD} and Appendix~\ref{app:fourth_moment}, respectively.
The steady state excess kurtosis in two dimension defined as
\bea
\mathcal{K}_{r}^{\rm st}=\f{\la\rv^4\ra_{r}^{\rm st}}{2 (\la\rv^2\ra_{r}^{\rm st})^2} - 1\,,
\eea
where $\la\rv^2\ra_{r}^{\rm st}$ is presented in Eq.~\eqref{eq:r2avg_reset_st} and $\la\rv^4\ra_{r}^{\rm st}$ is presented in Eq.~\eqref{eq:r4avg_reset_st}.
We discussed the steady-state excess kurtosis heatmap in the $r$–$\W_0$ plane, highlighting different states in Fig.~\ref{fig4}(a) of the main text.
Moreover, Figure~\ref{app_fig3} displays a heatmap of $\mathcal{K}_{r}^{\rm st}$ on the $D_r-\W_0$ plane revealing a re-entrant transition, from a resetting‑dominated regime to an active regime and back to resetting at low rotational noise strengths.

In the absence of translational noise ($D=0$), the state diagrams in the $r$–$\W_0$ and $D_r$–$\W_0$ planes (Fig.~\ref{app_fig_state_diagrams_D=0}(a) and Fig.~\ref{app_fig_state_diagrams_D=0}(b), respectively) show that the re-entrant behavior with chirality disappears. We refer to this limit, where translational noise is absent, as the chiral ABP without translational noise.
Furthermore, in the absence of both translational and rotational noise ($D = D_r = 0$), the state diagram in the $r$–$\W_0$ plane (Fig.~\ref{app_fig_state_diagrams_D=0_Dr=0}) shows that the re-entrant behavior with the resetting rate also disappears. We refer to this fully noiseless limit as the chiral active particle (CAP).

We expand in the small activity limit (letting active speed $v_0\to 0$)
\bea
\mathcal{K}_{r}^{\rm st} &=& 1 +\f{r[(r+D_r)^2-\W_0^2]v_0^2}{D[(r+D_r)^2+\W_0^2]^2} +\mathcal{O}(v_0^4)\,.
\eea
The analysis indicates that $\Omega_0$ is the critical parameter driving the second term negative (the reason behind the weakly active state), thereby suppressing the excess kurtosis.

\section{Calculation of $\la(\uv\cdot\rv)\rv^2\ra_s$}
\label{app:urr2}
We use $\psi=\la(\uv\cdot\rv)\rv^2\ra_s$ in Eq.~\eqref{eq:moment} to calculate the Laplace form of $\la(\uv\cdot\rv)\rv^2\ra_s$. We obtain $\la\psi\ra_0=0$, $\la\nabla^2\psi\ra_s=8\la\uv\cdot\rv\ra_s$, $\la\nabla_n^2\psi\ra_s=-\la(\uv\cdot\rv)\rv^2\ra_s$, $\la\uv\cdot\nabla\psi\ra_s=\la\rv^2\ra_s+2\la(\uv\cdot\rv)^2\ra_s$, $\la\rv\cdot\nabla\psi\ra_s=5\la(\uv\cdot\rv)\rv^2\ra_s$, and $\la\uv^{\perp}\cdot\nabla\psi\ra_s=\la(\uv^{\perp}\cdot\rv)\rv^2\ra_s$. Substituting this into the moment equation leads to
\bea
\la(\uv\cdot\rv)\rv^2\ra_s &=& \f{1}{s+D_r}\left[8D\la\uv\cdot\rv\ra_s+v_0\la\rv^2\ra_s+2 v_0\la(\uv\cdot\rv)^2\ra_s+\W_0 \la (\uv^{\perp}\cdot\rv)\rv^2\ra_s\right]\,,
\eea
where $\la\uv\cdot\rv\ra_s$ already calculated in Eq.~(\ref{eq:rn-Laplace}) and $\la\rv^2\ra_s$ in Eq.~(\ref{eq:r2avg-Laplace}). To proceed further we need to calculate $\la(\uv\cdot\rv)^2\ra_s$ and $\la (\uv^{\perp}\cdot\rv)\rv^2\ra_s$. First, we use $\psi=\la(\uv\cdot\rv)^2\ra_s$ in moment generator Eq.~(\ref{eq:moment}) 
\bea
\la(\uv\cdot\rv)^2\ra_s &=& \f{1}{s+4}\left[2D\la 1\ra_s+2\la\rv^2\ra_s+2 v_0\la\uv\cdot\rv\ra_s+2\W_0 \la(\uv\cdot\rv)(\uv^{\perp}\cdot\rv)\ra_s\right]\,,\\
\la(\uv\cdot\rv)(\uv^{\perp}\cdot\rv)\ra_s &=& \f{1}{s+4}\left[2\la\rv^2\ra_s+v_0\la\uv^{\perp}\cdot\rv\ra_s+\W_0 \la(\uv^{\perp}\cdot\rv)^2\ra_s-\W_0 \la(\uv\cdot\rv)^2\ra_s\right]\,.
\eea
Now, from the symmetry, $\la(\uv^{\perp}\cdot\rv)^2\ra_s = \la\rv^2\ra_s -\la(\uv\cdot\rv)^2\ra_s$,
\bea
\la(\uv\cdot\rv)(\uv^{\perp}\cdot\rv)\ra_s &=& \f{1}{s+4}\left[\W_0\la\rv^2\ra_s+v_0\la\uv^{\perp}\cdot\rv\ra_s-2\W_0 \la(\uv\cdot\rv)^2\ra_s\right]\,.
\eea
Proceed further,
\bea
\la(\uv\cdot\rv)^2\ra_s &=& \f{s+4}{(s+4)^2+(2\W_0)^2}\left[2D\la 1\ra_s+2\la\rv^2\ra_s+2 v_0\la\uv\cdot\rv\ra_s+\frac{2\W_0}{s+4} \left[\W_0\la\rv^2\ra_s+v_0\la\uv^{\perp}\cdot\rv\ra_s\right]\right]\,,\\
\la(\uv^{\perp}\cdot\rv)\rv^2\ra_s &=& \f{1}{s+D_r}\left[ 8D\la\uv^{\perp}\cdot\rv\ra_s+2 v_0\la(\uv\cdot\rv)(\uv^{\perp}\cdot\rv)\ra_s-\W_0 \la (\uv\cdot\rv)\rv^2\ra_s\right]\,.
\eea
Finally, we obtain $\la(\uv\cdot\rv)\rv^2\ra_s$,which is used in the derivation of the fourth order moment of displacement in Appendix~\ref{app:fourth_moment}; see Eq.~\eqref{eq:urr2}. 

\end{document}